\documentclass[12pt,nohyper]{JHEP3} 
\usepackage{epsfig}

\makeatletter
\def\art{\@ifnextchar[{\eart}{\oart}}
\def\eart[#1]#2#3#4#5#6{{\rm #2}, {#3 #4} {\rm (#6) #5} [arXiv:\-{{#1}}]}
\def\hepart[#1]#2{{\rm #2, arXiv:\-{#1}}}
\def\circa#1{\,\raise.3ex\hbox{$#1$\kern-.75em\lower1ex\hbox{$\sim$}}\,}

\title{Non-universal minimal $Z^\prime$ models: 
\\ present bounds and early LHC reach} 
\author{ Ennio Salvioni,$^{a}$  Alessandro Strumia,$^{b,c}$ Giovanni Villadoro$^{c}$ and Fabio Zwirner$^{a}$\\
$^a$ Dipartimento di Fisica, Universit\`a di Padova and INFN, Italy

$^b$ Dipartimento di Fisica, Universit\`a di Pisa and INFN, Italy

$^c$ Theory Group, Physics Department, CERN, Geneva, Switzerland

E-mail: \email{ennio.salvioni@pd.infn.it, Alessandro.Strumia@df.unipi.it, giovanni.villadoro@cern.ch, fabio.zwirner@pd.infn.it}
} 
\preprint{CERN-PH-TH/2009-212 \\ 
DFPD-09/TH/22} 
\abstract{
We consider non-universal `minimal' $Z^\prime$ models, whose additional $U(1)$ charge is a non-anomalous linear combination of the weak hypercharge $Y$, the baryon number $B$ and the partial lepton numbers $(L_e,L_\mu,L_\tau)$, with no exotic fermions beyond three standard families with right-handed neutrinos. We show that the observed pattern of neutrino masses and mixing can be fully reproduced by a gauge-invariant renormalizable Lagrangian, and flavor-changing neutral currents in the charged lepton sector are suppressed by a GIM mechanism. We then discuss the phenomenology of some benchmark models. The {\em electrophilic} $B-3 L_e$ model is significantly constrained by electroweak precision tests, but still allows to fit the hint of an excess observed by CDF in dielectrons but not in dimuons. The {\em muonphilic} $B- 3 L_\mu$ model is very mildly constrained by electroweak precision tests, so that even the very early phase of the LHC can explore significant areas of parameter space. We also discuss the {\em hadrophobic} $L_\mu - L_\tau$ model, which has recently attracted interest in connection with some puzzling features of cosmic ray spectra.}
\begin{document} 

\section{Introduction} 
\label{intro}

Extensions of the Standard Model (SM) at the TeV scale, with an additional $U(1)$ factor in the gauge group associated with a heavy neutral gauge boson $Z^\prime$, have often been considered, with various theoretical motivations (for recent reviews and references, see e.g. \cite{reviews}). 

In a recent paper \cite{svz}, three of us discussed the phenomenology of {\em minimal} $Z'$  models, previously introduced in \cite{adh} and identified as the most economical U(1) extensions of the SM that do not spoil renormalizability. Making reference to the SM particle content, the key ingredients of {\em minimality} are: no exotic vectors, apart from a single $Z^\prime$ associated with a $U(1)$ factor in the gauge group, commuting with $G_{SM} = SU(3)_C \times SU(2)_L \times U(1)_Y$; no exotic fermions, apart from one right-handed neutrino, singlet under $G_{SM}$, for each of the three SM families; anomaly cancellation.

An additional simplifying assumption made in \cite{svz}  was the  family-independence of the additional $U(1)$ charge, which could then be identified with an arbitrary linear combination of the weak hypercharge $Y$ and $B-L$, where $L=L_e+L_\mu+L_\tau$ is the total lepton number. In such a case, electroweak precision tests (EWPT) strongly constrain the parameter space of the models, leaving little virgin land to be explored in the very early phase of the LHC, when energy and luminosity will be limited \cite{earlylhc} with respect to the design parameters. 

In this paper, we relax the assumption made in \cite{svz}, allowing the additional $U(1)$ generator to be an arbitrary non-anomalous linear combination of the weak hypercharge $Y$, the baryon number $B$ and the partial lepton numbers $L_e,L_\mu,L_\tau$. It is immediate to check (see, e.g., \cite{nonuniv}) that, when baryon and lepton numbers appear only in a linear combination $X$ of the three generators $(B-3 L_a)$ ($a=e,\mu,\tau$), any such model is anomaly-free. Indeed, it is easy to check that, if we add the requirement that all fermion masses and mixing be generated by a renormalizable gauge-invariant Lagrangian, with a suitable Higgs content, then this is the most general solution. Models have already been considered with gauged $B - 3 L_\tau$ \cite{btau}, $B - 3 L_\mu$ \cite{bmu} and $L_a-L_b$ ($a \ne b$) \cite{lilj}. In particular, models with gauged $B - 3 L_\mu$ were suggested\footnote{One of us (A.S.) thanks C.~Boehm  for discussions about why, after correcting a sign mistake, $B-3 L_\mu$ does not fit the NuTeV anomaly \cite{CB}.} as possible non-SM explanations \cite{bmu}  of the NuTeV `anomaly' \cite{NuTeV}, and models with gauged $L_a - L_b$  ($a \ne b$) as non-SM explanations \cite{liljpam} of possible excesses in $e^\pm$ cosmic ray spectra \cite{Pamela}. Some constraints on the above models from EWPT were discussed in \cite{lebed}, and other anomaly free models, where neutrino masses are generated by non-renormalizable gauge-invariant interactions, were considered in \cite{nonuniv}.

Our paper is organized as follows. In Section~\ref{theory}, we discuss the theoretical input for the following phenomenological analysis. We first review the structure of the fermionic neutral currents. We then show that the observed pattern of fermion masses and mixings, in particular the one in the neutrino sector, can be obtained  from a gauge-invariant renormalizable Lagrangian, with no more fine-tuning of the Yukawa couplings than the one already needed in the SM to reproduce the observed electron mass. Moreover, all flavor-changing neutral current processes (FCNC) involving charged leptons are strongly suppressed by the neutrino mass differences, as in the Glashow-Iliopoulos-Maiani (GIM) mechanism \cite{GIM} of the SM. We end this theoretical section by studying the renormalization group equations (RGE) for the effective couplings controlling the weak neutral currents, as well as the $Z$--$Z'$ mixing. Assuming that the model is valid up to very high scales, such as some super-unification or grand-unification scale, we identify a favored region for the effective low-energy couplings and mixing. In Section~\ref{pheno} we review the general aspects of the phenomenology of the class of models under consideration. We start with the constraints from electroweak precision tests, where we complete and update the fit of \cite{CCMS}, based in turn on \cite{bprs} and used in \cite{svz} for constraining the universal models where $X=B-L$. We then briefly recall the procedure followed in \cite{svz} for extracting the Tevatron bounds and assessing the discovery prospects for the very early phase of the LHC, to be repeated here for non-universal minimal $Z'$ models. In Section~\ref{models}  we apply our results to the study of three benchmark models that we find particularly interesting: 
\begin{enumerate}
\item
an {\em electrophilic} model, corresponding to $X=B - 3 L_e$, which could explain, for values of the parameters not excluded by EWPT, a small excess observed at the Tevatron in the dielectron sample \cite{CDFepem,D0epem}, at invariant masses around 240~GeV, but not in the dimuon sample \cite{CDFmumu};  
\item
a {\em muonphilic} model,  corresponding to $X=B - 3 L_\mu$, which is subject to much milder constraints from EWPT, thus has a much wider area of parameter space accessible, via the dimuon signature, already in the very early phase of the LHC. In a sense, this model is another example of the {\em supermodels} recently discussed in \cite{supmod}, defined as those for which the LHC sensitivity with only 10~pb$^{-1}$ is greater than that of the Tevatron with 10~fb$^{-1}$; 
\item
a {\em hadrophobic} model, corresponding to $X = L_\mu - L_\tau$, which has recently attracted considerable interest \cite{liljpam} as a possible non-SM explanation of the positron excess in cosmic ray data \cite{Pamela}.
\end{enumerate}
Finally, in Sect.~\ref{concl} we present our conclusions.

\section{Theory} 
\label{theory}

\subsection{Parameterization} 
\label{param}

Along the lines of \cite{svz}, we work in a field basis where gauge boson kinetic terms are canonical and gauge boson masses are diagonal, and write the two fermionic currents coupled to the neutral massive gauge bosons as:
\begin{equation}
J_Z^\mu  =  \cos \theta^\prime \, J_{Z^0}^\mu -  \sin \theta^\prime \, J_{Z^{\prime \, 0}}^\mu  \, , 
\qquad
J_{Z^\prime}^\mu  =  \sin \theta^\prime \, J_{Z^0}^\mu +  \cos \theta^\prime \, J_{Z^{\prime \, 0}}^\mu  \, , 
\label{currents}
\end{equation}
where
\begin{equation}
J_{Z^0}^\mu =g_Z   \sum_f   \, \overline{f}  \, \gamma^\mu \left( T_{3L} - \sin^2 \theta_W \, Q \right)  \, f  \, , 
\qquad
\left( g_Z = \sqrt{g^2+g^{\prime \, 2}} \right) \, , 
\label{zcurr}
\end{equation}
is the SM expression for the current coupled to the SM $Z^0$  (we recall that, in the presence of mixing, $Z^0$ does not coincide with the mass eigenstate $Z$), and
\begin{equation}
 J_{Z^{\prime \, 0}}^\mu = \sum_f \overline{f} \, \gamma^\mu \left( g_Y \, Y + g_X \, X \right) \,  f  =g_{Z} \sum_f   \, \overline{f} \, \gamma^\mu \, Q_{Z'} \,  f   \, . 
\label{zpcurr}
\end{equation}
In the above expressions, $f=(u_{La},d_{La},u_{Ra},d_{Ra},\nu_{La},e_{La},\nu_{Ra},e_{Ra})$ runs over the different chiral projections of three families of SM fermions, $Q =T_{3L} + Y$ is the electric charge and
\begin{equation}
\label{xdef}
X = \sum_{a =e , \mu, \tau} \frac{\lambda_a}{3}  \, (B - 3 L_a) = \beta B-\lambda_e L_e-\lambda_\mu L_\mu-\lambda_\tau L_\tau\,,
\qquad  \beta \equiv \frac13\sum_a \lambda_a \,
\end{equation}
where the $\lambda_a$ are three arbitrary real coefficients. The charges of the SM fermions needed for evaluating the currents of Eqs.~(\ref{currents})--(\ref{zpcurr}) are collected\footnote{Notice that, with respect to \cite{svz}, we switch from a basis of left-handed fermions and antifermions to a basis of left- and right-handed fermions.} in Tab.~\ref{charges}, where for expressing $Q_{Z'}$ we make use of the ratios 
\begin{equation}
\label{ratios}
\widetilde{g}_Y \equiv \frac{g_Y}{g_Z} \, , 
\qquad
\widetilde{g}_X \equiv \frac{g_X}{g_Z} \, . 
\end{equation}
\TABLE[t]{
\begin{tabular} { | c | c | c | c | c | c | }
\hline
& $T_{3L}$ & $Y$ & $B - 3 L_b$ & $X$ & $Q_{Z'}$  \\
\hline
&&&&& \\
$q_{La} \equiv \left( \begin{array}{c} u_L \\ d_L \end{array} \right)_a$ 
&
$\left( \begin{array}{c} +\frac{1}{2} \\ -\frac{1}{2} \end{array} \right)$ 
&
$+ \frac{1}{6}$
&
$+ \frac{1}{3}$
&
$+ \frac{\beta}{3}$
&
$\frac{1}{6} \, \widetilde g_Y  + \frac{\beta}{3} \, \widetilde g_X $
\\
&&&&& \\
\hline
&&&&& \\
$u_{Ra}$ 
&
0
&
$+ \frac{2}{3}$
&
$+ \frac{1}{3}$
&
$+ \frac{\beta}{3}$
&
$\frac{2}{3} \, \widetilde g_Y  + \frac{\beta}{3} \, \widetilde g_X$
\\
&&&&& \\
\hline
&&&&& \\
$d_{Ra}$  
&
0
&
$- \frac{1}{3}$
&
$+ \frac{1}{3}$
&
$+ \frac{\beta}{3}$
&
$- \frac{1}{3} \, \widetilde g_Y  + \frac{\beta}{3} \, \widetilde g_X$
\\
&&&&& \\
\hline
&&&&& \\
$ l_{La} \equiv \left( \begin{array}{c} \nu_L \\ e_L \end{array} \right)_a$ 
& 
$\left( \begin{array}{c} +\frac{1}{2} \\ -\frac{1}{2} \end{array} \right)$ 
&
$- \frac{1}{2}$
&
$-3 \, \delta_{ab} $
&
$- \delta_{ab} \, \lambda_b$
&
$-\frac{1}{2} \, \widetilde g_Y - \delta_{ab} \, \lambda_b \, \widetilde g_X$
\\
&&&&& \\
\hline
&&&&& \\
$\nu_{Ra}$
& 
0
&
0
&
$- 3 \, \delta_{ab} $
&
$- \delta_{ab} \, \lambda_b$
&
$- \delta_{ab} \, \lambda_b \, \widetilde g_X$
\\
&&&&& \\
\hline
&&&&& \\
$e_{Ra}$ 
&
0
&
$-1$
&
$- 3 \, \delta_{ab} $
&
$- \delta_{ab} \, \lambda_b$
& 
$- \widetilde g_Y - \delta_{ab} \, \lambda_b \, \widetilde g_X$ \\
&&&&&\\
\hline
\end{tabular}
\caption{The charges of the SM fermions controlling the weak neutral currents.}
\label{charges}
}

Since the SM Higgs doublet $H$ has\footnote{This property is shared by the MSSM Higgs doublets $H_1$ and $H_2$.} $X = 0$, and for breaking $X$ we consider only $SU(2)_L$-singlet Higgs fields, which must have $Y = Q - T_{3L} =0$, to avoid the spontaneous breaking of the electric charge, we can express the $Z$-$Z^\prime$ mixing angle $\theta^\prime$ in terms of $g_Y$ and $M_{Z^\prime}$,
\begin{equation}
\label{mixing}
\tan \theta^\prime = - \widetilde{g}_Y \, \frac{M_{Z^0}^2}{M_{Z^\prime}^2 - M_{Z^0}^2} \, ,
\qquad\hbox{where}\qquad
M_{Z^0}^2 = \frac{g_Z^2 \, v^2}{4}
\end{equation}
is the SM expression for the $Z^0$ mass. Notice that $\theta' \neq 0$ whenever $\widetilde g_Y\neq0$, because in such a case gauge invariance of the SM Yukawa terms forces the SM Higgs to be charged under the extra $U(1)$, thus producing a $Z$-$Z'$ mixing.

We can then study the $Z^\prime$ phenomenology in terms of five unknown parameters: 
\begin{itemize}
\item the $Z^\prime$ mass $M_{Z^\prime}$ and the two coupling constants $(g_Y,g_X)$ or, equivalently, $(\widetilde g_Y,\widetilde g_X)$;
\item two independent ratios of the three flavor $\lambda_a$ coefficients (the overall scale can be reabsorbed in $g_X$) that define $X$ in Eq.~(\ref{xdef}).
\end{itemize}
We will mention later possible additional parameters describing the enlarged Higgs sector and the right-handed neutrino masses, but we will mostly consider physical contexts in which these additional parameters play a negligible r\^ole.

\subsection{Fermion masses and mixing} 
\label{neum}

To make the model realistic, we have to make sure that the observed pattern of fermion masses and mixing can be generated. Our philosophy is to stick to renormalizable interactions, formally sending the mass scale of non-renormalizable operators to infinity, and to introduce the minimal number of Higgs fields that can realize the required gauge and flavor breaking. 

We gauge baryon number $B$ rather than baryon flavors, so that, in the quark sector, masses and mixing can be generated exactly as in the SM, by
\begin{equation}
\label{qyuk}
- {\cal L}_{\rm Yuk}^{(q)} = \overline{u_R} \, Y^U \, q_L H +
\overline{d_R} \, Y^D \, q_L \widetilde{H} + {\rm h.c.} \, , 
\end{equation}
where obvious contractions of $SU(3)$, $SU(2)$ and family indices are understood, 
\begin{equation}
\label{smhiggs}
H = \left( \begin{array}{c} H^+ \\ H^0 \end{array} \right)  \sim (+1/2,0) \, , 
\qquad
\widetilde{H} = i \, \sigma^2 \, H^\dagger =  \left( \begin{array}{c} H^0 \\ - H^- \end{array} \right)  \sim (-1/2,0)  
\end{equation}
are the SM Higgs field and its conjugate (the numbers in brackets are the charges $Y$ and $X$, respectively), $Y^U$ and $Y^D$ are $3 \times 3$ complex matrices, unrestricted by the additional $U(1)$ gauge invariance. As in the SM, the transition from the interaction basis to the mass basis for the quark fields generates the CKM matrix but preserves the diagonal form of the quark contribution to all tree-level gauge neutral currents. 

We then write down the most general Yukawa couplings involving the SM Higgs in the leptonic sector. Their general form will be as in the SM,
\begin{equation}
\label{lyuk}
- {\cal L}_{\rm Yuk}^{(l)} = \overline{e_R} \, Y^E \, l_L \widetilde{H} +
\overline{\nu_R} \, Y^N \, l_L H + {\rm h.c.} \, , 
\end{equation}
but, since we gauge lepton flavor rather than lepton number,  some of the entries in the complex $3 \times 3$ matrices $Y^E$ and $Y^N$ can be now forbidden by the extra $U(1)$. To understand this point, we can treat $Y^E$ and $Y^N$ as spurions, and write down the $X$ charges of their individual entries:
\begin{equation}
\label{yukx}
X(Y^E_{ab}) = X(Y^N_{ab}) = \lambda_b - \lambda_a \, .
\end{equation}
Therefore, diagonal entries are always allowed, whilst those off-diagonal entries for which $\lambda_b \neq \lambda_a$  are forbidden.  Anyway, there is no obstruction to the generation of the observed values of the charged lepton masses. 

For neutrino masses, we choose to make use of the (type-I) see-saw mechanism, to avoid fine-tuning the Yukawa couplings in $Y^N$ to a higher extent than those in $Y^E$. We then allow for right-handed neutrino mass terms of the form
\begin{equation}
\label{massr}
{\cal L}_M^{(\nu)} = \frac{1}{2} \, \overline{(\nu_R)} \, M_R (\varphi) \, \overline{\nu_R}^T + {\rm h.c.} \, , 
\end{equation}
where $M_R(\varphi)$ is a complex symmetric $3 \times 3$ matrix that may contain bare mass terms as well as terms proportional to non-SM Higgs fields, singlets under the SM gauge group but carrying a non-vanishing $X$ charge. As before, we can treat the entries of $M_R(\varphi)$ as spurions and write:
\begin{equation}
\label{massx}
X[M_R(\varphi)_{ab}] =  \lambda_a + \lambda_b \, .
\end{equation}
For entries with $X=0$, bare (field-independent) neutrino mass terms are allowed, whose size is not restricted by gauge invariance and unitarity. For entries with $X \ne 0$, mass terms can be generated by introducing renormalizable Yukawa couplings to the Higgs(es) $\varphi_X$ that break the extra U(1), with charge $X$ under the extra U(1) and neutral under the SM gauge group:
\begin{equation}
\label{bsmhiggs}
\varphi_X \sim (0, X) \, , 
\end{equation}
in the same notation of Eq.~(\ref{smhiggs}). Therefore, depending on their $X$ charges, some right-handed neutrinos might acquire mass only after the breaking of the extra U(1). In conclusion, the right-handed neutrino masses can be of order $M_{Z'}$ or higher.
The three heavy neutrinos can be made sufficiently heavy to become negligible in the discussion of the $Z'$ phenomenology. Similarly, the physical degrees of freedom in the non-SM Higgs fields of Eq.~(\ref{bsmhiggs}), whose vacuum expectation values (vevs) give mass to the $Z'$ and to right-handed neutrinos, can be made sufficiently heavy and/or sufficiently decoupled from the SM Higgs field to be also negligible in the present discussion of the $Z'$ phenomenology.  
 
We conclude this section by showing that our minimal non-universal renormalizable\footnote{Here we differ from \cite{nonuniv}, which aimed at forbidding neutrino mass terms from renormalizable couplings, in the attempt of explaining in such a way the smallness of neutrino masses. We insist instead, coherently with the requirement of anomaly cancellation, on the renormalizabiliy of the theory at the TeV scale, meaning that all non-renormalizable operators allowed by the symmetries are suppressed by a very high physical cut-off scale. We do not see a problem, in the present context, in having the Yukawa couplings in $Y^N$ of the same of order of the electron Yukawa coupling in $Y^E$.} $Z'$ models are compatible with the observed neutrino masses and mixing (for a review, see, e.g., \cite{strvis}), which turn out to contain, as in the SM, all lepton-flavor violation of experimental interest: the new $Z'$ gauge interactions do not lead to dangerous FCNC processes involving charged leptons, which get suppressed by a GIM-like mechanism \cite{GIM}.  We want to reach the usual basis where the charged lepton mass matrix is diagonal and the mass matrix for the three light neutrinos is written in terms of the mass eigenvalues $m_{1,2,3}$  and the mixing matrix $U$, as
\begin{equation}
m^\nu = U^* \cdot  {\rm diag} (m_1,m_2,m_3) \cdot U^\dagger \, .
\end{equation}
So, we first diagonalize the charged lepton mass matrix,
\begin{equation}
M^E = Y^E \, \langle H^0 \rangle \, ,
\end{equation}
by performing the usual unitary transformations
\begin{equation}
l_L \rightarrow U_L \, l_L \, , 
\qquad
e_R \rightarrow U_R \, e_R \, ,
\end{equation}
where both $U_L$ and $U_R$ are $3 \times 3$ unitary matrices, so that
\begin{equation}
U_R^\dagger \, M^E \, U_L = {\rm diag}(m_e,m_\mu,m_\tau) \, ,
\end{equation}
and we can identify the redefined $\nu_L$ as the neutrino flavor eigenstates, partners of the charged lepton mass and flavor eigenstates in $SU(2)_L$ doublets. Correspondingly, the Yukawa couplings associated to Dirac neutrino masses get redefined as $Y^N \rightarrow Y^N \, U_L^\dagger$.  After the rotation the coupling of the $Z'$ to the leptons reads
\begin{equation}
g_Z Z'_\mu \left ({\overline {l_L}} \gamma^\mu U_L^\dagger Q_{Z'} U_L  l_L+{\overline {e_R}} \gamma^\mu U_R^\dagger Q_{Z'} U_R  e_R+ {\overline {\nu_R}} \gamma^\mu  Q_{Z'}  \nu_R \right )\,.
\end{equation}
The important observation is  that $U_L$ and $U_R$ do not mix  sectors with different $X$ charges, so that flavor mixing is present only between flavors with the same $X$ charge, thus with the same $Z'$ interaction, and no tree-level FCNC involving the SM charged leptons are generated. Flavour changing neutral currents do appear in the neutrino sector, as a consequence of the rotation required for diagonalizing the neutrino mass matrix. Such FCNC are however suppressed by the light neutrino masses, as in the SM, and do not produce dangerous effects.

We now show that this restricted framework, leaving generic the right-handed neutrino mass matrix $M_R$,  has enough freedom to reproduce the observed light neutrino masses and mixing. The standard see-saw formula
\begin{equation}
m^\nu = (M^N)^T \cdot M_R^{-1} \cdot  M^N \, , 
\end{equation}
%
where $M^N = Y^N \langle H^0 \rangle$, can be trivially solved for $M_R=(M^N)^T \cdot (m^\nu)^{-1}\cdot (M^N)$,  when $\det m^\nu \ne 0$, meaning that any\footnote{The argument remains of course true in the case of one massless neutrino, still allowed by the data.}  neutrino mass matrix $m^\nu$ can be obtained from any given Dirac mass matrix $M^N$ for an appropriate $M_R$.

\subsection{Renormalization group effects} 
\label{sec:rge}

We recall that we denoted as $g$ and $g'$ the SM gauge couplings of $SU(2)_L \times U(1)_Y$, and as $g_X$ and $g_Y$ the gauge couplings of the $X$ and $Y$ components  of the extra U(1) current, see Eq.~(\ref{zpcurr}). A given U(1) is specified by assigning $X$, see Eq.~(\ref{xdef}). The values of these couplings are scale-dependent, and their scale-dependence is controlled by the corresponding RGE. In the formalism of \cite{svz}, but referring this time directly to the three $U(1)$ gauge coupling constants, we can write at one-loop:
\begin{eqnarray}
\label{rgeX}
\frac{dg_X}{dt} & =& \frac{1}{16 \pi^2} \left( b_{XX} \, g_X^{3}
+  2 \, b_{YX} \, g_{X}^{2} \, g_{Y} + b_{YY} \, g_{X} \, g_{Y}^{2} \right) \,,\\
\label{rgepr}
\frac{dg'}{dt} & =& \frac{1}{16 \pi^2} \,  b_{YY} \, {g'}^{3}\,,\\
\label{rgeY}
\frac{dg_{Y}}{dt} & =&  \frac{1}{16 \pi^2} \left(  b_{YY} \, g_{Y} \, (g_{Y}^{2}+2{g'}^{2}) + 2 \, b_{YX} \, g_{X} \, (g_{Y}^{2}+{g'}^{2}) + b_{XX} \, g_{X}^{2} \, g_{Y} \right) \, , 
\end{eqnarray}
where $t= \log (Q/Q_0)$, $Q_0$ is a reference scale, 
\begin{equation}
\label{bab}
b_{AB} = \frac{2}{3} \sum_f Q_f^A Q_f^B + \frac{1}{3} \sum_s Q_s^A Q_s^B \, ,
\qquad
(A,B=Y,X) \, , 
\end{equation}
and $f$ and $s$ are the two-component fermions and the complex scalars in the theory, respectively.

In the following, when discussing our benchmark models, we will identify some GUT-favored regions in the $(\widetilde{g}_Y,\widetilde{g}_X)$ plane according to the following procedure. For choosing the boundary conditions at the GUT scale $M_U$, we normalize all $U(1)$ charges as $T_{3L}$, over three fermion generations with right-handed neutrinos, and we take $M_U = 10^{16} \, {\rm GeV}$ as a reference value. In typical GUTs, $M_U$  can vary within approximately two decades around such reference value, but the difference in our estimate of the GUT-favored region is of the order of other threshold effects that we reabsorb in the wide ranges we assume below for other parameters. Then, we compute the boundary value $g^\prime (M_U)$ using the phenomenological input $g^\prime (M_Z) = e(M_Z)/\cos \theta_W (M_Z)$, with $\alpha_{em}^{-1}(M_Z) \simeq 128$ and $\sin^2 \theta_W (M_Z) \simeq 0.23$, and the SM one-loop RGE. We then allow the $Z'$ coupling at the unification scale $\alpha_U = g_U^2/(4 \pi) = g_{Z'}^2 (M_U) /(4 \pi)$, to vary within the generous range
\begin{equation}
\label{aurange}
\frac{1}{100} \circa{<} \alpha_U \circa{<} \frac{1}{20} \, . 
\end{equation}
Taking into account that the SM RGE would predict $\alpha_U \sim 1/45$, our upper and lower bounds leave a margin of more than a factor of two to account for threshold corrections, new particles at the TeV scale and other model-dependent effects. Correspondingly, we determine the GUT-favored region of the $(\widetilde{g}_Y, \widetilde{g}_X)$ plane by making use of the one-loop RGE of eqs.~(\ref{rgeX})-(\ref{bab}): when discussing different benchmark models, the result will be presented as a colored band. In addition, we will study the stability with respect to the RGE evolution of the `pure--$X$' models, in which $g_Y = 0$ at the unification scale $M_U$.

%
%

\section{Phenomenology} 
\label{pheno}

\subsection{Electroweak precision tests} 
\label{ewpt}

As already mentioned, EWPT can be used to constrain the parameter space of minimal $Z'$ models. The analysis of \cite{svz} was limited to the {\em universal} case where $X=B-L$, and was done under some approximations that allowed to make use of the results of \cite{CCMS} without performing a new fit. In particular: instead of the full set of relevant electroweak precision data, only nine pseudo-observables were used; the experimental and theoretical inputs of the fit were  not updated beyond Winter 2006; the constraints coming from the muon anomalous magnetic moment (for a review, see, e.g., \cite{g-2}) and from the NuTeV experiment \cite{NuTeV} on neutrino-hadron deep inelastic scattering were not included. 

Here we perform a new, full global fit, employing the most recent measurements of the following precision observables performed at LEP1, Tevatron, SLC and other facilities: pole masses $M_Z$, $M_W = (80.399\pm0.023)\,{\rm GeV}$~\cite{newwmass} and $m_t = (173.1 \pm 1.3) \, {\rm GeV}$~\cite{mt}; Fermi constant for $\mu$ decay; strong and electromagnetic coupling at $M_Z$; total $Z$ width; $e^+e^-$ hadronic cross section at the $Z$ peak, forward-backward asymmetry in the $\ell$ $(=e,\mu,\tau)$, $b$, $c$ final states; $\tau$ polarization asymmetry; $Z$ branching fraction into hadrons, $b\bar{b}$, $c\bar{c}$; left-right polarization asymmetry in the $\ell$ $(=e,\mu,\tau)$, $b$, $c$ final states.

We also include the following LEP2 data, measured at various values of $\sqrt{s}$ between 183 and 207 GeV: $\sigma(e^+e^- \to q \bar q)$; cross-sections and forward-backward asymmetries for $e^+e^-\to \mu^+\mu^-$, $\tau^+\tau^-$, $b\bar b$; angular distribution $d\sigma(e^+e^-\to e^+e^-)/d\cos\theta$ as measured by OPAL and ALEPH.

We additionally include the following low-energy measurements, which are less relevant in the fit to universal models, but can play a r\^ole in some region of the parameter space of non-universal models: M\o{}ller scattering at $Q^2=0.026\,{\rm GeV}^2$; atomic parity violation (APV) in Cs as re-analyzed in~\cite{apvth}:  $Q_W = -73.16\pm0.35$; $g_L$ and $g_R$ in neutrino-nucleon scattering as measured by NuTeV; anomalous magnetic moment of the muon~\cite{g-2}. The last three observables are plagued by controversial theoretical uncertainties: we adopt the standard estimates and will warn the reader whenever these less relevant measurements play a significant r\^ole in the global fit.

In the above lists, we provided numerical values and references only for those observables that have been updated with respect to~\cite{CCMS, bprs}. 

The corrections to all the above observables are computed by integrating out the heavy gauge boson coupled to the total (fermionic and Higgs) $J^\mu_{Z^{\prime \, 0}}$ current, obtaining the effective Lagrangian:
\begin{equation}
{\cal L}_{\rm eff} = {\cal L}_{\rm SM} - \frac{(J_{Z^{\prime \, 0}})^\mu (J_{Z^{\prime \, 0}})_\mu}{2 \, M_{Z'}^2} \, ,
\end{equation}
valid under the assumption that we are perturbing around the SM fit. The term proportional to $|H^\dagger iD_\mu H|^2/M_{Z'}^2$ affects the $Z$ mass, whilst the squared fermion current gives four-fermions operators. The mixed term affects the $Z$ couplings to fermions.

The results of our improved and updated fit, applied to the parameter space of the minimal universal $Z'$ model with $X=B-L$, discussed in \cite{svz}, are displayed in Fig.~\ref{fig:ewptBL}:
\FIGURE[t] {
\includegraphics[width=0.49\textwidth]{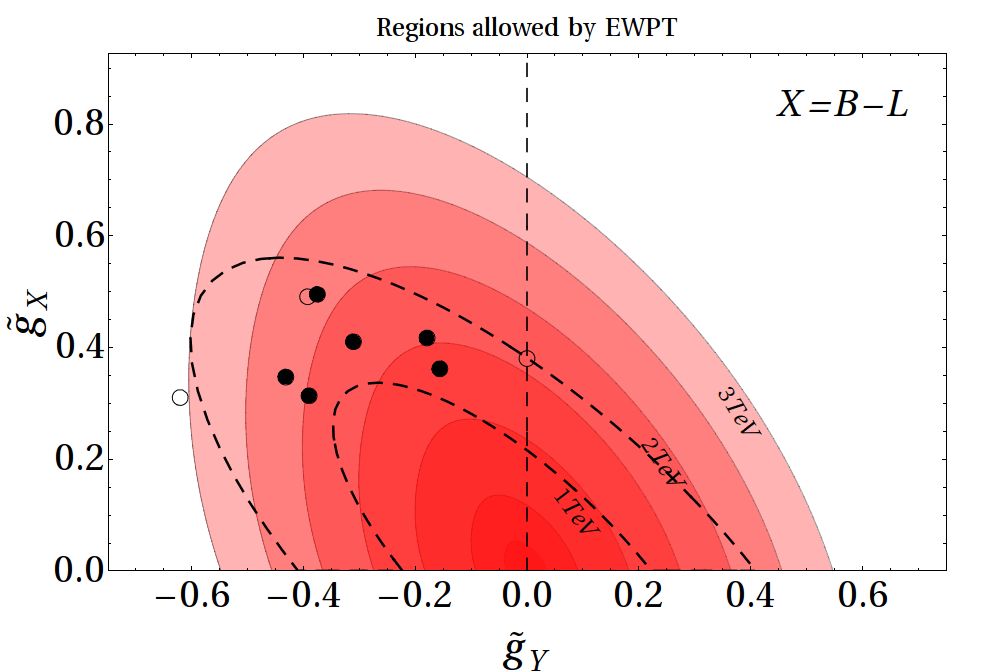}
\includegraphics[width=0.49\textwidth]{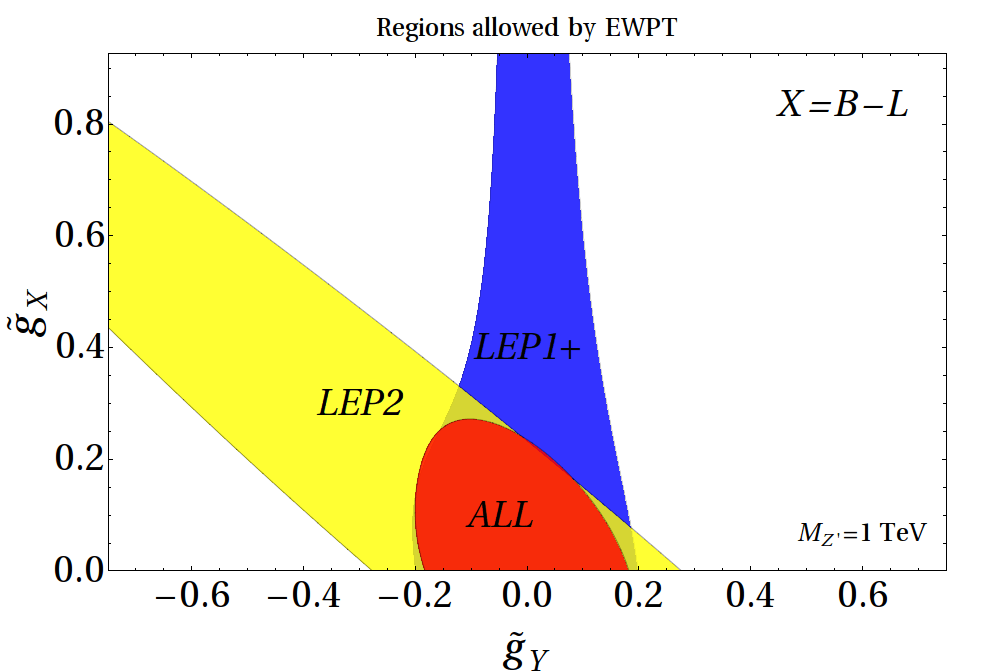}
\caption{Universal model, $X$$=$$B$$-$$L$. \emph{Left:}  The regions allowed by all  EWPT at 95\% CL, for $M_{Z'} = 200$, $500$, $1000$, $1500$, $2000$, $2500$ and $3000$~GeV (from inner to outer). \emph{Right:} The regions allowed at 95\% CL, for $M_{Z'}=1$~TeV, by LEP2 (yellow), LEP1-APV-NuTeV-SLD (blue)  and all combined (red).}
\label{fig:ewptBL}}
for some representative values of the $Z'$ mass, the corresponding 95\% CL allowed regions in  the $(\widetilde{g}_Y, \widetilde{g}_X)$ plane are shown. On the left-hand side, the region allowed by the global fit is the red one enclosed by each contour, and the different mass values are (from inner to outer): $M_{Z'} = 200$, $500$, $1000$, $1500$, $2000$, $2500$ and $3000$~GeV. Dashed lines and empty (full) dots remind us of the GUT-favored region and of the GUT-inspired (-derived) benchmark models, whose precise definitions can be found in \cite{svz}. 

Comparing with the corresponding figure of \cite{svz}, we see that our improved and updated fit gives slightly more stringent bounds, and that the ellipsoidal curves
delimiting the allowed regions are now less tilted with respect to the vertical axis. A detailed check shows that the effect is mostly due to the transition from an approximate to a full fit, and is in agreement with the findings of \cite{CCMS}. The updates of some input data such as $m_t$ and $M_W$, and the inclusion of some new ones such $(g-2)_\mu$ and NuTeV, do not affect the result in a significant way. To understand better the different effects, the right-hand side of the figure shows, for the representative value   $M_{Z'} = 1$~TeV, how the regions allowed by LEP2 (yellow) and LEP1-APV-NuTeV-SLD (LEP1+ for short, blue) combine to give the globally allowed (red) region. The behavior is the expected one: data at the Z peak and at lower energies\footnote{We are dealing here with a vector-like charge $X$ for the extra $U(1)$.} are mostly sensitive to mixing effects, thus the blue band surrounds the vertical axis that corresponds to vanishing mixing; data at LEP2 energies are mostly sensitive to $Z'$-exchange diagrams with some external electrons, thus the axis of the yellow band roughly corresponds to the direction that minimizes the $Z'$ couplings to electrons.   

As in \cite{svz}, we report in Tab.~\ref{tab-mzbounds} the bounds on the $Z'$ masses (in a 1-parameter fit) for some particular choices of the couplings corresponding to GUT-inspired and SUSY-GUT benchmark models. 
\TABLE[t] {
\begin{tabular}{|c||c|c|c|c|c|c|c|c|c|} 
\hline 
$\phantom{\sqrt{\frac11}}$ &${Z'}_{B-L}^{(0)}$ & ${Z'}_{B-L}^{\rm (iii)}$ &${Z'}_{B-L}^{\rm (iv)}$ 
& ${Z'}_\chi^{(0)}$   &${Z'}_\chi^{\rm (iii)}$ &${Z'}_\chi^{\rm (iv)}$
&${Z'}_{3R}^{(0)}$&${Z'}_{3R}^{\rm (iii)}$   & ${Z'}_{3R}^{\rm (iv)}$ \\ \hline \hline
$M_{Z'}$ (TeV) & 1.80 & 1.77 & 1.53 & 2.61 & 2.54 & 2.11 & 3.64 & 2.61 & 2.36 \\ \hline
\end{tabular}
\label{tab-mzbounds}
\caption{95\% CL bounds (1-parameter fit) on the $Z'$ masses from EWPT, corresponding to the specific universal models ($X=B-L$) represented by the nine points in Fig.~\ref{fig:ewptBL}. The ${Z'}^{(0)}$ models are those represented by empty points, while ${Z'}^{\rm(iii)}$ (${Z'}^{\rm(iv)}$) corresponds to the three external (internal) black points; see \cite{svz} for details.}
}
We observe that, because of the slightly different inclination of the ellipsoidal curves with respect to those from the approximate fit of \cite{svz}, the bounds on the `$B-L$ models' are now slightly less stringent, whilst the bounds on the `$\chi$-models' and on the `$T_{3R}$-models' are now slightly more stringent.

The validity of our fit to EWPT extends to the non-universal minimal models considered in this paper: the corresponding bounds on the parameter space will be illustrated in Sect.~\ref{models} for the three chosen benchmark models. 

Before concluding this discussion, we would like to point out that the bounds obtained in our fit are consistent with those given in \cite{carena}, but considerably more stringent than those obtained by other authors (see, e.g., \cite{reviews}). We identify two specific reasons for this difference. First, some authors do not take the constraints from LEP2 into full account; the right-hand side of Fig.~\ref{fig:ewptBL} shows that LEP2 data are important. Second, some authors treat the $Z$-$Z'$ mixing angle $\theta^\prime$ as an independent free parameter, the $\chi^2$-minimization then selects a special value of $\theta^\prime$ very close to $0$, which strongly relaxes the bounds from EWPT. However, for any given model, the value of $\theta^\prime$ is fixed by the gauge couplings and by the Higgs vevs to a value that, in general, is different from zero: in our class of models, it is fixed by Eq.~(\ref{mixing}). In models with a more exotic Higgs content, the cancellation of the mixing angle can be achieved only by fine-tuning the gauge couplings and the Higgs vevs and would be scale-dependent. Without such fine-tuning, also in those models the $Z'$ bounds from EWPT would become much stronger, of the order of those reported in Tab.~\ref{tab-mzbounds}.
 
\subsection{Direct searches at hadron colliders} 
\label{tevatron}

We briefly recall here the general procedure followed in \cite{svz} for extracting the bounds from direct searches at the Tevatron and for assessing the discovery prospects in the very early phase of the LHC, to be repeated here for non-universal minimal $Z'$ models. This will prepare the ground for Section~\ref{models}, where we will apply such a procedure to the study of our three benchmark models and comment on their peculiar features.
 
Both at the Tevatron and at the LHC, we concentrate on the processes $p \bar p \ (pp) \to (Z' \to \ell^+ \ell^-) + X$, ($\ell = e , \mu$): these are two very clean channels to look for, and after simple generous cuts the irreducible background is dominated by the well-understood SM Drell-Yan (DY) processes. For the class of models under consideration, other channels into SM final states are not expected to be competitive for exclusion or discovery, even if they could play a r\^ole in the determination of the $Z^\prime$ couplings after a future discovery.

For any given model in our class, specified by the choice of $X$, we compute the Tevatron $Z'$ production cross-section multiplied by the branching ratio into two charged leptons, $\sigma(p\bar p\to Z'X) \times {\rm BR}(Z'\to \ell^+ \ell^-)$, as a function of the three parameters $M_{Z'}$, $g_Y$, $g_X$), and compare it with the limits established by the CDF and D0 experiments.

On the theory side, we perform the calculation at NLO in QCD (and LO for the EW part), using the NLO MSTW08 PDF sets \cite{mstw08}. In the calculation of the total width $\Gamma_{Z'}$  we include the following channels: $Z'\to f\bar f$, $W^+W^-$, and $Zh$,  where $h$ is the SM Higgs boson and $f$ are the SM fermions of Tab.~\ref{charges},  with the exception of the right-handed neutrinos, which we take to be heavier than $M_{Z'}/2$. The ratio $\Gamma_{Z^\prime} / M_{Z'}$ is pretty constant over the whole range of masses of interest, and is of order a few percent for GUT-favored $Z'$ couplings, and of course smaller for more weakly coupled $Z'$. 

For the Tevatron experimental limits, we use the most recent available results from CDF (on $Z'\to e^+e^-$~\cite{CDFepem} and $Z'\to\mu^+\mu^-$~\cite{CDFmumu}) and D0 (on $Z'\to e^+e^-$~\cite{D0epem}).  They directly provide the 95\% CL bounds on the product $\sigma(p\bar p\to Z^\prime \, X)\times {\rm BR}(Z^\prime \to \ell^+\ell^-)$ based on $2.5$, $2.3$, $3.6$~fb$^{-1}$ of data, with $27\div38$\%, $13\div40$\%, $17\div22$\% total acceptances respectively, growing from smaller to larger values of $M_{Z'}$. Notice that, although D0 data refer to a higher integrated luminosity, the acceptance is smaller, making the D0 bounds a little weaker than those from CDF.

When compared with the computed cross-section, the experimental limits produce, for each value of $M_{Z'}$, a 95\% CL exclusion region in the  $(\widetilde g_Y, \widetilde g_X)$ plane, in analogy with the EWPT case. 

The other question we want to address is the following. At what combined values of center-of-mass (CoM) energy and integrated luminosity may we expect the LHC to start having a chance of discovering a $Z'$ (at least of the kind discussed in this work), taking into account all the experimental bounds from EWPT and Tevatron direct searches? What region of parameter space that has not been already ruled out could be accessible for different luminosities and energies in the first LHC runs? Considering the fact that $Z'$ signals are among the cleanest and easiest ones in the search of new physics, our analysis may also be used as a benchmark point when discussing the integrated luminosities that are worth collecting at each energy to actually probe new physics.

At the moment, the program for the first year of LHC running \cite{earlylhc} consists in a first run at low energy ($\sqrt{s} = 7 \, {\rm TeV}$) and low luminosity ($<100$~pb$^{-1}$), followed by an upgrade in energy ($\sqrt{s} \leq 10 \, {\rm TeV}$), with a collected luminosity up to 200$\div$300~pb$^{-1}$. At such low\footnote{Of course, with respect to the LHC design parameters.} energies and luminosities, the constraints from Tevatron direct searches and EWPT will play a crucial r\^ole in identifying the allowed region of parameter space that can be probed and the time scale required to have access to it.

Being the LHC a hadron collider, the region of parameter space accessible to the LHC is similar in shape to the corresponding one at the Tevatron. For relatively light $Z'$ ($M_{Z^\prime}  < 800$~GeV), since the strongest constraints come from Tevatron direct searches, we expect the LHC to turn into a discovery machine as soon as it becomes sensitive to regions of parameters not yet excluded by the Tevatron. However, while the higher energy is clearly a big advantage for intermediate $Z'$ masses of several hundreds GeV, for lighter masses the low luminosity may be a crucial limiting factor in the early LHC phase. For heavier $Z'$ masses, such as those relevant for GUT models with $X=B-L$, the discussion is more subtle than in the case of universal models discussed in \cite{svz}. If the $Z'$ couples significantly to electrons, then generically EWPT outperform the Tevatron, and the LHC must wait for higher energies and luminosities to become sensitive. If, on the other hand, the $Z'$ couples to muons much more than to electrons, then the constraints from EWPT are considerably weaker, and the LHC has a significant discovery potential already at a very early phase.  

To turn these considerations into more quantitative statements, we perform a basic analysis along the lines of the one described before for extracting the Tevatron bounds. In the present case we consider the range $\sqrt{s}=7\div10$~TeV for the $pp$ CoM energy, luminosities in the range $50$~pb$^{-1}\div1$~fb$^{-1}$, and calculate the product $\sigma (pp \to Z'X) \times {\rm BR} (Z'\to \ell^+\ell^-)$ for $M_{Z'}=200 \div 3000$~GeV, at the same order in perturbation theory as in the Tevatron case.
At the same level of precision, we also compute the SM Drell-Yan (DY) differential cross-section, which constitutes the main source of background.

To gain some approximate understanding of the acceptances for signal and background at different values of the invariant mass $M_{\ell^+\ell^-}$ of the $\ell^+ \ell^-$ pair, and of the possible model-dependence of the former, we performed a simple study, scanning over different non-universal minimal models: its conclusions are essentially the same as for the universal minimal models of \cite{svz}. Since our computed values of the acceptance are compatible with those of ref.~\cite{cmsmu}, we adopt their Fig.~2 for the rest of our LHC study.  More refined studies, however, could take into account also the model-dependence of the acceptance, which may not be negligible for $Z'$ searches at relatively small masses.

To estimate the 5$\sigma$ discovery reach of the early phase of the LHC \cite{cmsmu,atlas,cmse}, we compare the events due to a generic $Z'$ signal to the events from the SM-DY background in a 3$\%$ interval around the relevant values of the dilepton invariant mass\footnote{This conservative assumption is compatible with the expected energy resolution, even in this early phase, and with the fact that, for GUT-favored values of the coupling constants, $\Gamma_{Z'} / M_{Z'} \sim {\rm few} \%$.}. We then require the signal events to be at least a $5\sigma$ fluctuation over the expected background, and in any case more than 3 events. This rough statistical analysis is enough to get an approximate answer to the questions we want to address. We leave a more careful analysis to the experimental collaborations ATLAS and CMS,  which have control on all the information needed to perform it in an accurate and reliable way. 


\section{Benchmark models} 
\label{models}

\FIGURE[t] {
\includegraphics[width=0.49\textwidth]{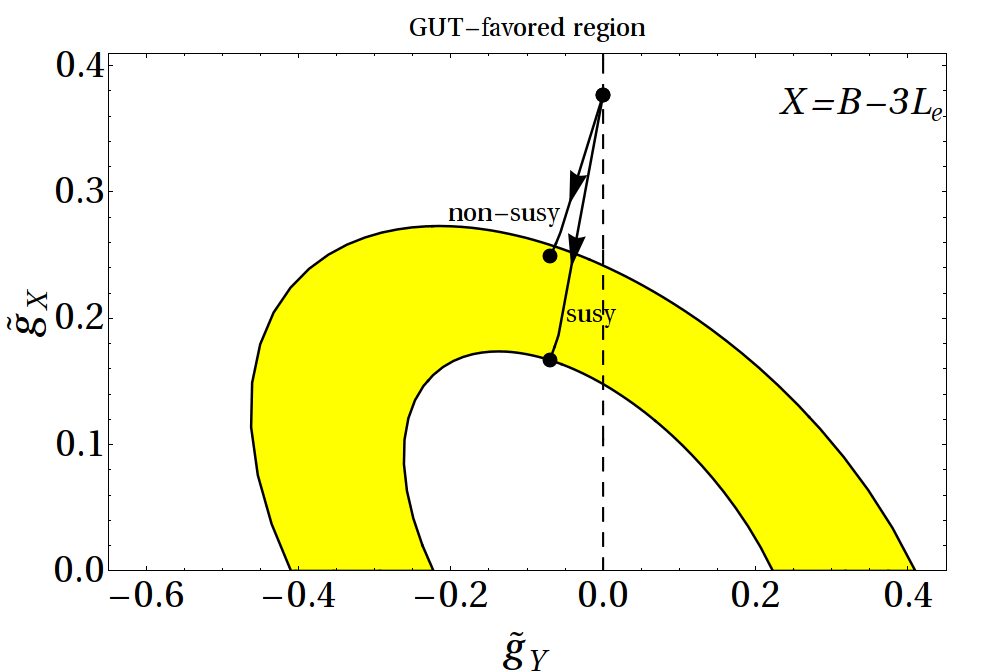}
\includegraphics[width=0.49\textwidth]{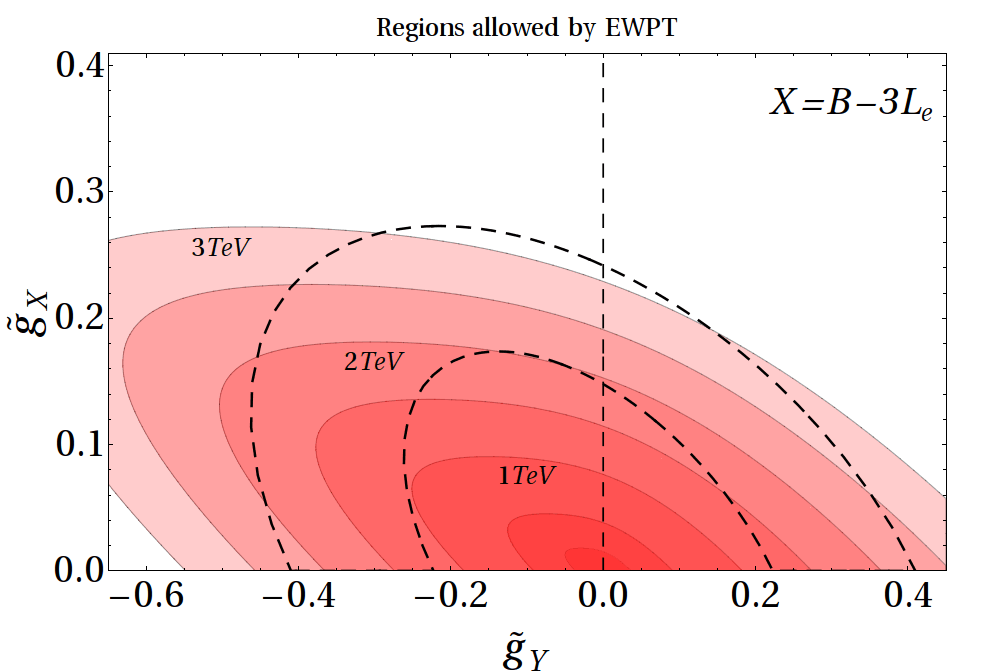}
\caption{Electrophilic model, $X$$=$$B$$-$$3 L_e$. \emph{Left:} The GUT-favored region in the $(\widetilde g_Y, \widetilde g_X)$ plane and the effect of RGE running from the boundary conditions $\alpha_U(10^{16}~{\rm GeV})=1/24$ and $\tilde g_Y(10^{16}~{\rm GeV})=0$ for the two cases (i) and (ii) discussed in the text (higher and lower black curves). \emph{Right:} The shaded regions are allowed at 95\% CL by EWPT for $M_{Z'}=200$, $500$, $1000$, $1500$, $2000$, $2500$ and $3000$~GeV (from inner to outer).}
\label{fig:gut-ewptBLe}}
\FIGURE[t]{
\includegraphics[width=0.49\textwidth]{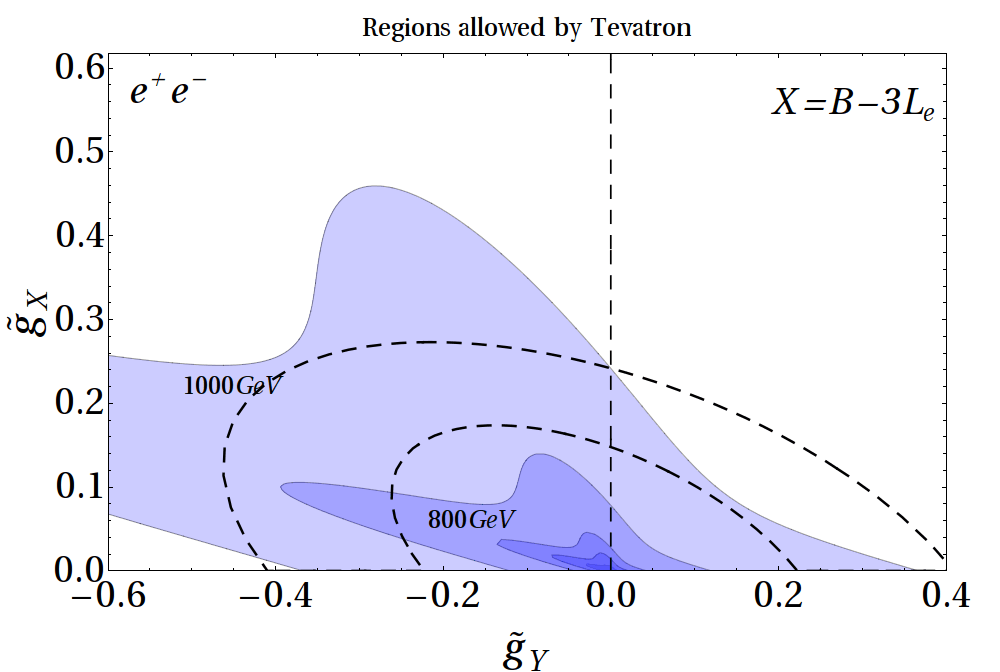}
\includegraphics[width=0.49\textwidth]{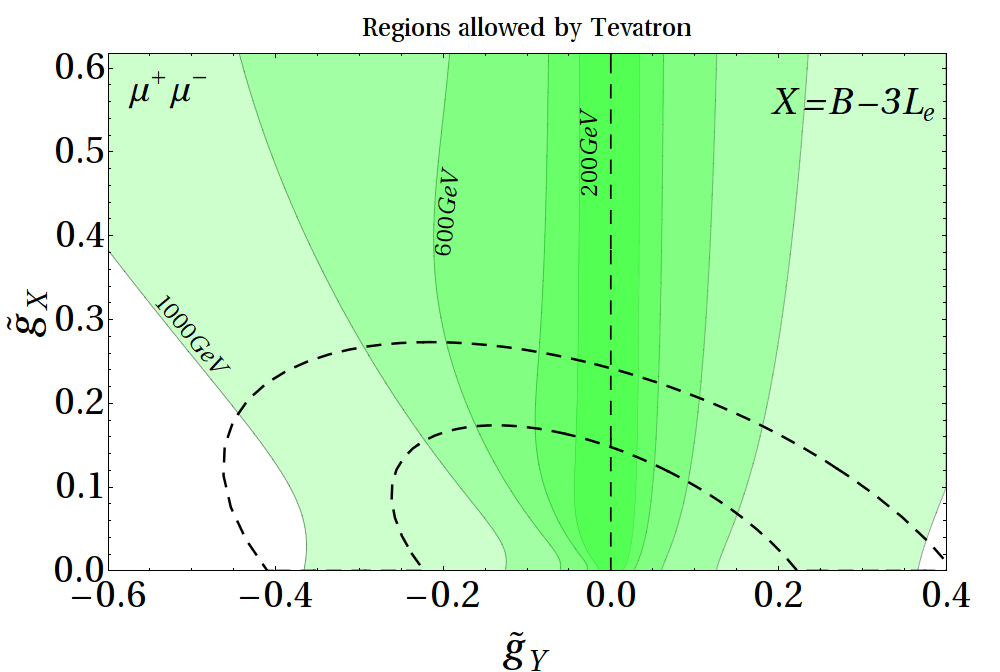}
\caption{Electrophilic model, $X$$=$$B$$-$$3 L_e$. The shaded regions are allowed at 95\% CL by dielectron  (\emph{left}) and dimuon (\emph{right}) searches at the Tevatron for $M_{Z'}=200$, $400$, $600$, $800$ and $1000$~GeV (from inner to outer).}
\label{fig:tev3BLe}}

\subsection{Electrophilic model: \boldmath{$X = B - 3 L_e$}} 
\label{electrophilic}

The first model we consider [corresponding to $\lambda_e=3$, $\lambda_\mu = \lambda_\tau = 0$ and $\beta = 1$ in Eq.~(\ref{xdef})] is associated to the gauging of a linear combination of $X=B-3L_e$ and the weak hypercharge $Y$. In this case, when the $Z'$ is `pure $B-3L_e$', it couples only to quarks and to first-generation leptons; in particular, the couplings to muons and taus only appear when the mixing with $Y$, parametrized by the effective coupling $g_Y$, is non-vanishing. For several aspects this case is similar to the universal case with $X=B-L$, with the exceptions of a stronger coupling to electrons than quarks and a suppression of the couplings to muons and taus. The construction of explicit GUT models, with a symmetry breaking chain that leads to the present example, would deserve a separate study, which goes beyond the aim of the present paper. Since we are not aware of any fundamental obstruction, we again identify, for illustrative purposes, a GUT-favored region in the $({\widetilde g}_Y,\widetilde g_X)$ plane, performing the RGE evolution from $M_U \sim 10^{16}$ down to the weak scale, with boundary conditions varied in the interval (\ref{aurange}). The coefficients of the beta functions in this case are:
\begin{center}
\begin{tabular} { l | c   c   c  }
model & $b_{YY}$ &$b_{YX}$ &$b_{XX}$ \\
\hline
\phantom{i}i) non-susy & 41/6 &16/3 &125/3 \\
ii) susy & 11 & 8  &130 
\end{tabular}
\end{center}
\FIGURE[t]{
\includegraphics[width=0.49\textwidth]{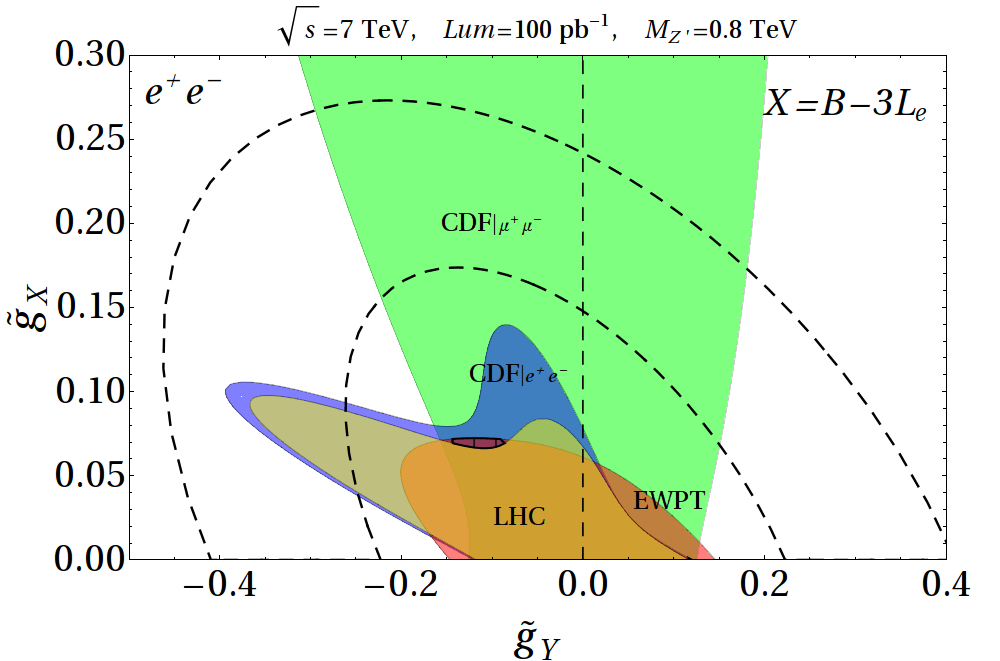}
\includegraphics[width=0.49\textwidth]{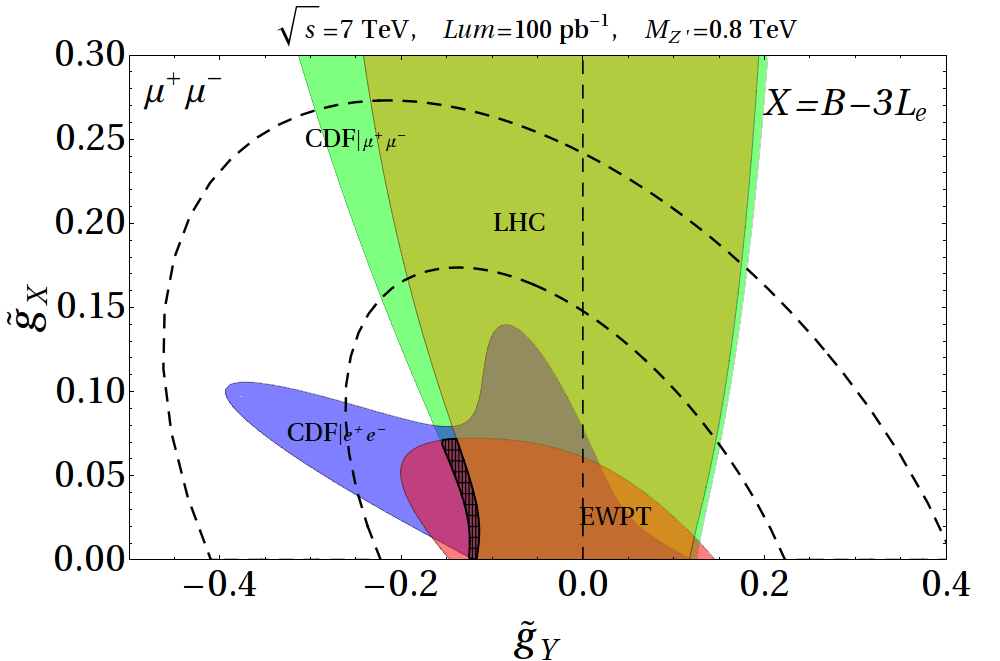}
\includegraphics[width=0.49\textwidth]{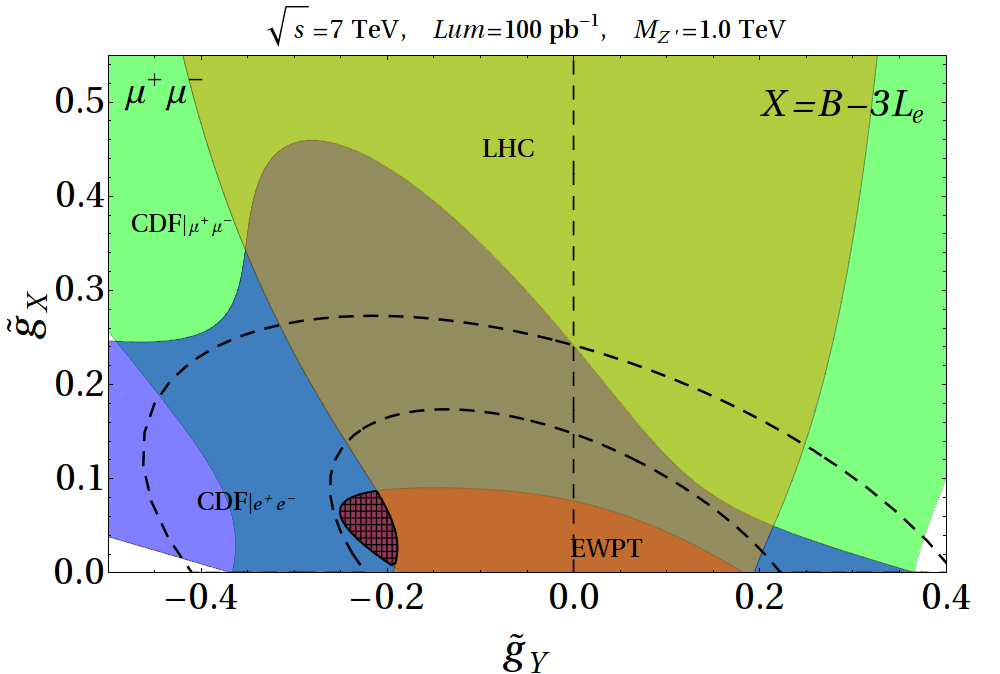}
\includegraphics[width=0.49\textwidth]{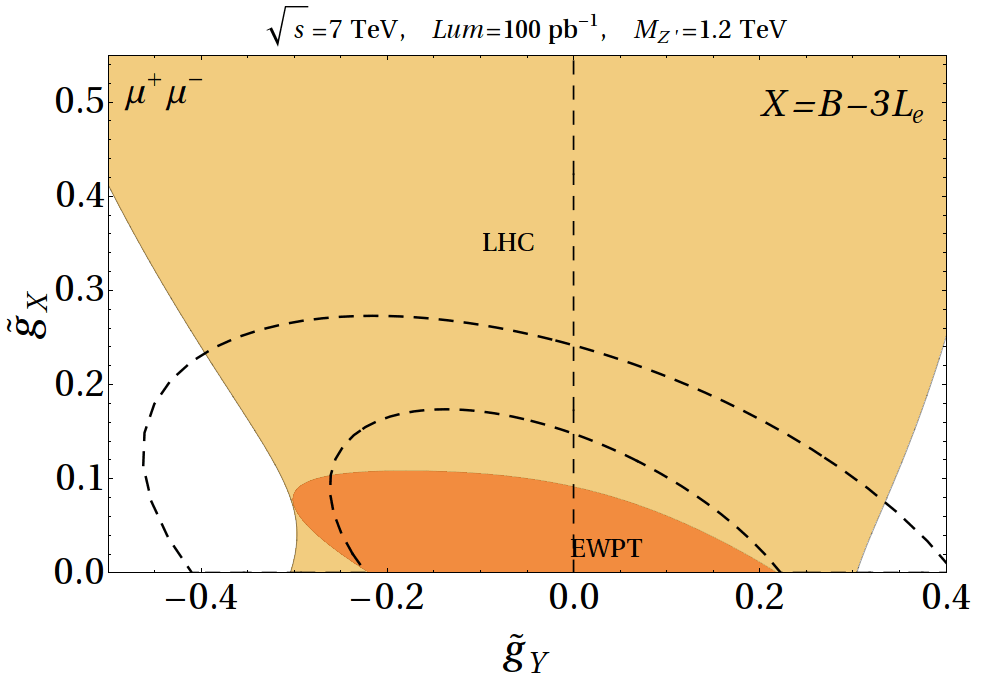}
\caption{Electrophilic model, $X$$=$$B$$-$$3L_e$. Regions allowed at $95\%$ by precision tests (red), by Tevatron $e^+ e^-$ data (blue), by Tevatron $\mu^+\mu^-$ data (green), compared with the region not accessible at the LHC (yellow) with $\sqrt{s}=7$~TeV, 100~pb$^{-1}$ of integrated luminosity for $M_{Z'} =$800, 800, 1000, 1200~GeV respectively. Regions are not shown when they would fill the whole plot. The first figure refers to dielectrons searches at the LHC, the other three to dimuon searches. The allowed region where the LHC can make a $5\sigma$ discovery is textured.}
\label{fig:lhcBLe7}}
where: model (i) corresponds to the SM field content plus right-handed neutrinos and two scalar Higgs fields, singlets under the SM gauge group, with charges $L_e=1$ and $L_e=2$; model (ii) corresponds to the MSSM field content plus the scalar superfields of three right-handed neutrinos and four Higgs bosons with charges $L_e = \pm 1$ and $L_e = \pm 2$. The result is plotted on the left-hand side of Fig.~\ref{fig:gut-ewptBLe}, which also shows the effect of the running (with the generated kinetic mixing) in the case we start from the representative values $\alpha_U(M_U)=1/24$ and $\widetilde g_Y(M_U)=0$ in both cases (i) and (ii). The shape of the preferred region is similar to the $X=B-L$ case.

The bounds from EWPT can be plotted on the same plane for different choices of the $Z'$ mass, see the right-hand side of Fig.~\ref{fig:gut-ewptBLe}.  Except for a small squeeze in the shape of the ellipsoidal curves representing  the bounds, induced by the change in the relative couplings between electrons and quarks, the bounds are similar to those for the universal $X=B-L$ case, since in EWPT most of the observables do not involve muons or taus.

The absence of couplings to muons in $Z'_X$ shows up more evidently in the bounds from direct searches at the Tevatron, see Fig.~\ref{fig:tev3BLe}. Indeed, while the $e^+e^-$ channel shows a roughly isotropic constraint, independent of the $X$-$Y$ nature of the $Z'$, as expected, the $\mu^+\mu^-$ channel shows no constraints for a $Z'$ that is pure $B-3L_e$. Notice that, on the contrary, when $B-3L_e$ mixes with $Y$, the $\mu^+\mu^-$ channel can produce stronger bounds than the $e^+e^-$ one; this effect is due to the fact (see Table~{charges}) that the $B-3L_e$ and $Y$
contributions to $Q_{Z'}$ for the electrons tend to cancel each other in the region $\tilde g_Y\sim -(3\div 6) \tilde g_X$ (the lower spikes on the left-hand side of Fig.~\ref{fig:tev3BLe}), this cancellation being absent for muons that have only the $Y$ contribution to $Q_{Z'}$.  

The shapes of the regions accessible to the LHC are identical to those excluded by the Tevatron, only the size is different, depending on the available luminosity and CoM energy. The results for the early phase are similar to the $B-L$ case: Fig.~\ref{fig:lhcBLe7} shows the accessible regions for different masses and channels after 100~pb$^{-1}$ of integrated luminosity at 7~TeV, while 
Fig.~\ref{fig:lhcBLe10} refers to 200~pb$^{-1}$ at 10~TeV. As for the $B-L$ case, with the early low-energy low-luminosity run, only very tiny regions of parameter space are available for discovery. The situation improves a bit only in the region of couplings smaller than those preferred by unification, after the first runs at higher energies. An interesting feature is that, at variance with flavor-universal models such as the one based on $X=B-L$, the two channels probe complementary regions of parameter space.
\FIGURE[t]{
\includegraphics[width=0.49\textwidth,height=0.18\textheight]{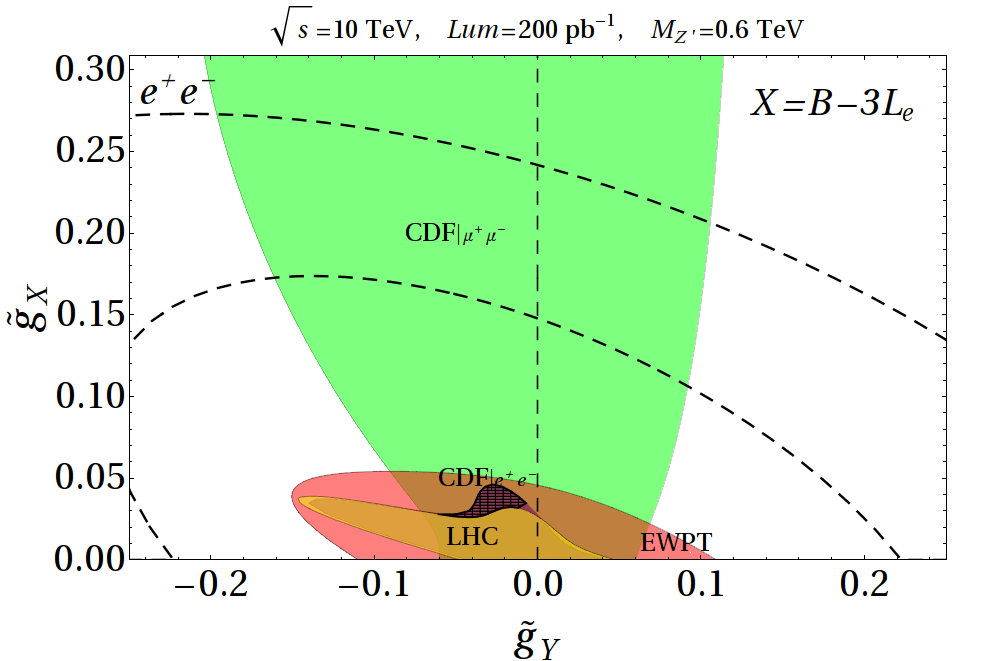}
\includegraphics[width=0.49\textwidth,height=0.18\textheight]{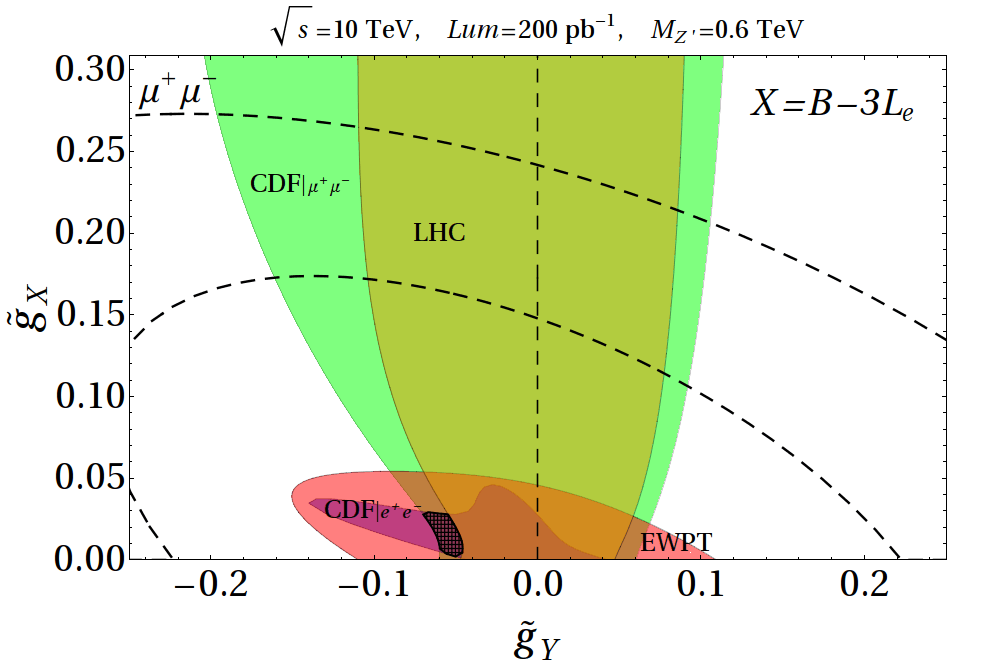}
\includegraphics[width=0.49\textwidth,height=0.18\textheight]{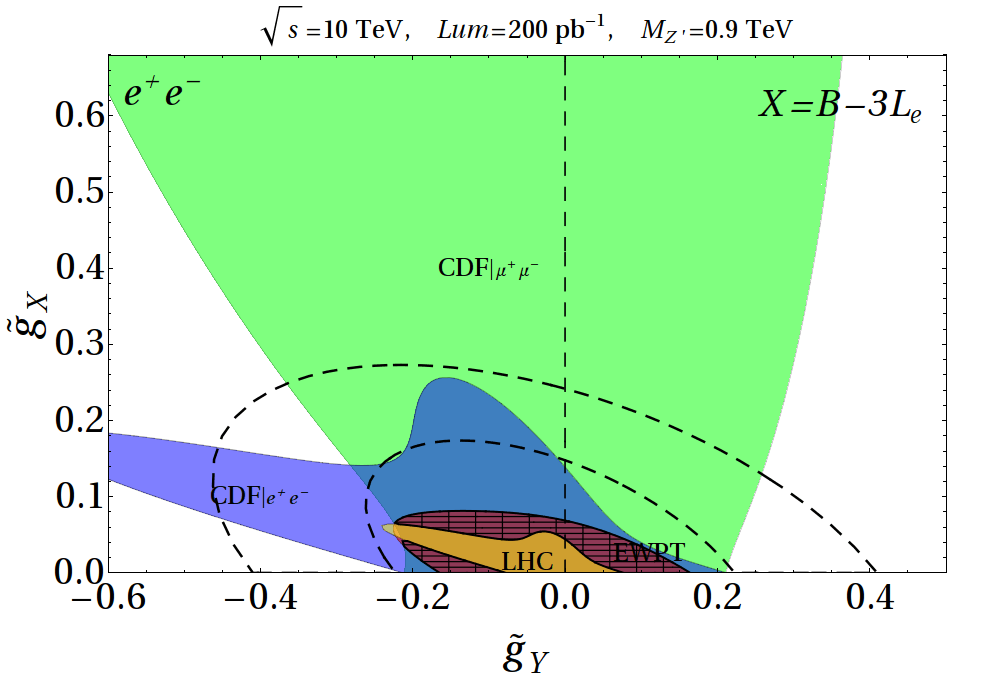}
\includegraphics[width=0.49\textwidth,height=0.18\textheight]{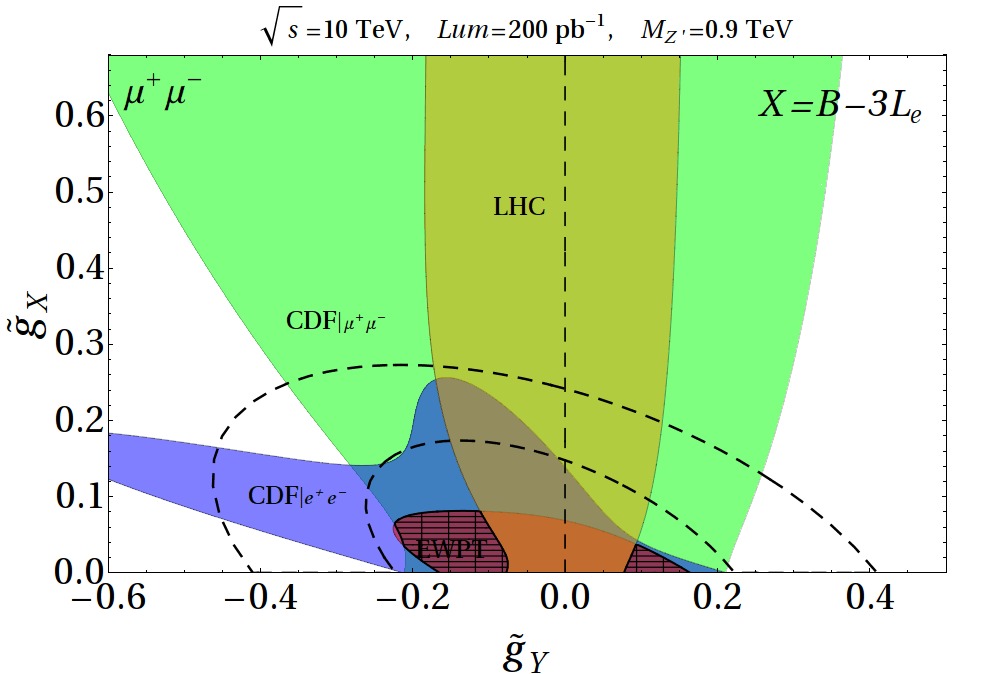}
\includegraphics[width=0.49\textwidth,height=0.18\textheight]{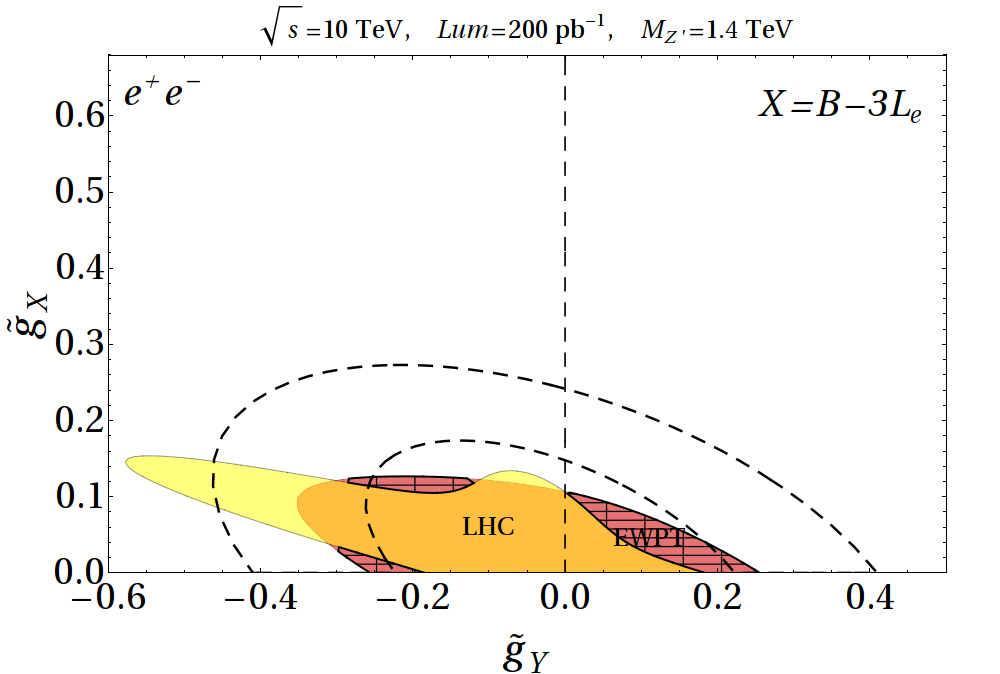}
\includegraphics[width=0.49\textwidth,height=0.18\textheight]{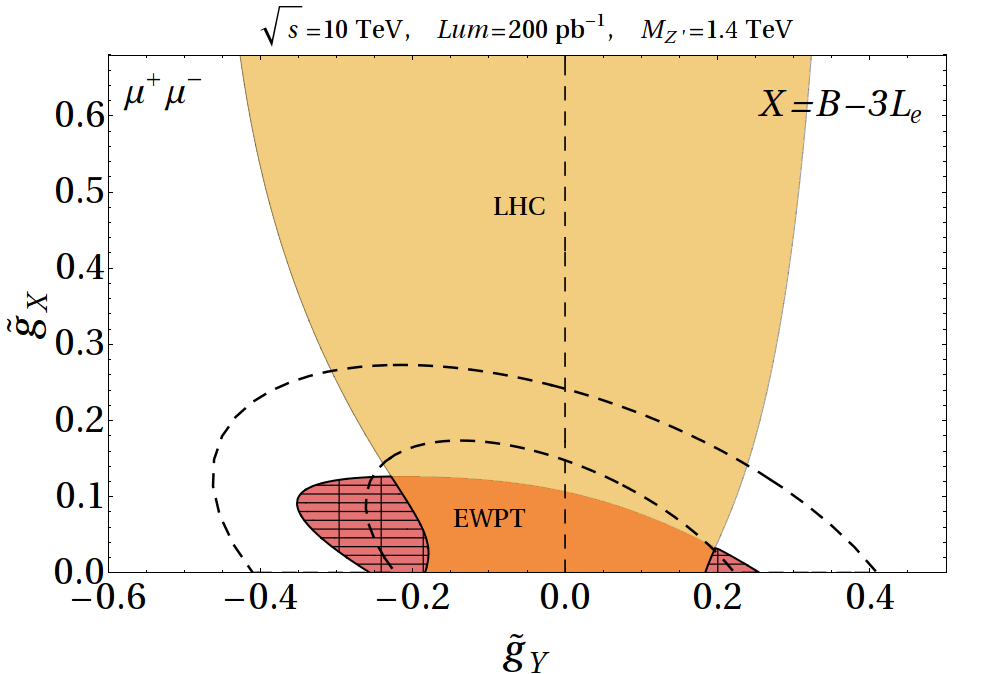}
\includegraphics[width=0.49\textwidth,height=0.18\textheight]{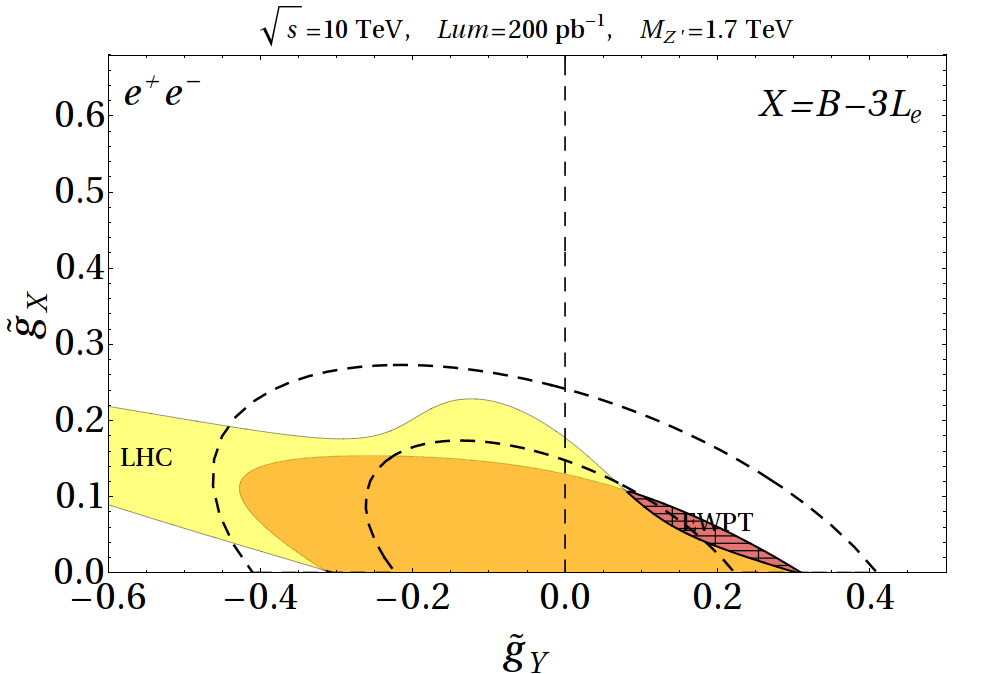}
\includegraphics[width=0.49\textwidth,height=0.18\textheight]{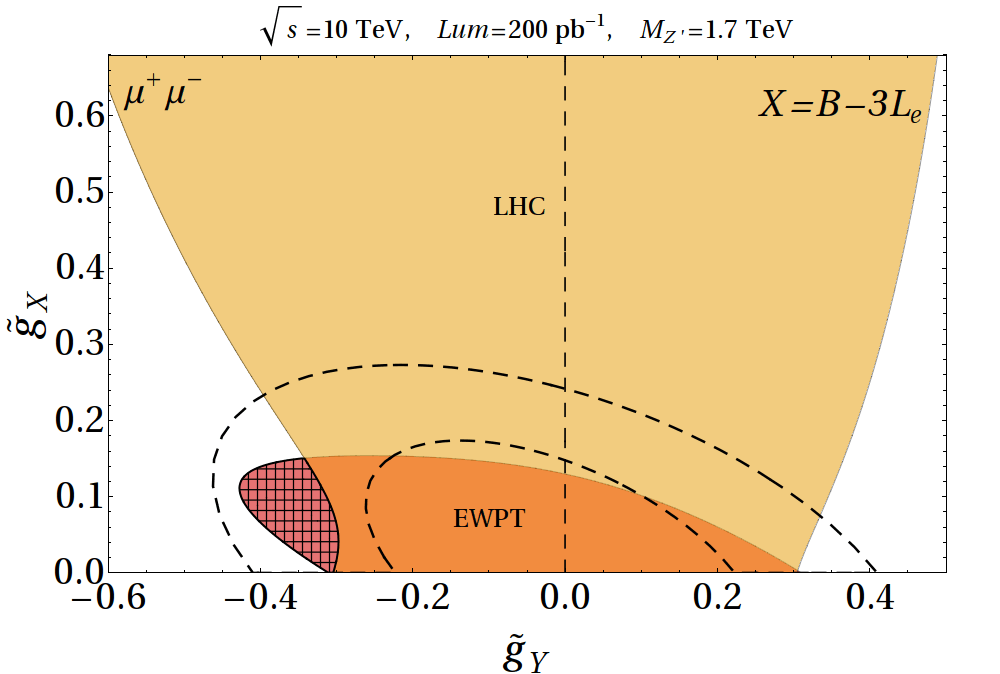}
\caption{Electrophilic model, $X$$=$$B$$-$$3L_e$. As in the previous figure, but for a better LHC with $\sqrt{s}=10$~TeV and 200~pb$^{-1}$ of integrated luminosity. The left (right) columns refers to $e^+e^-$ ($\mu^+\mu^-$) searches. The $Z'$ mass in the different rows is 600, 900, 1400, 1700~GeV respectively. The allowed region where the LHC can make a $5\sigma$ discovery is textured.}
\label{fig:lhcBLe10}
}
\FIGURE[t]{
\includegraphics[width=0.6\textwidth]{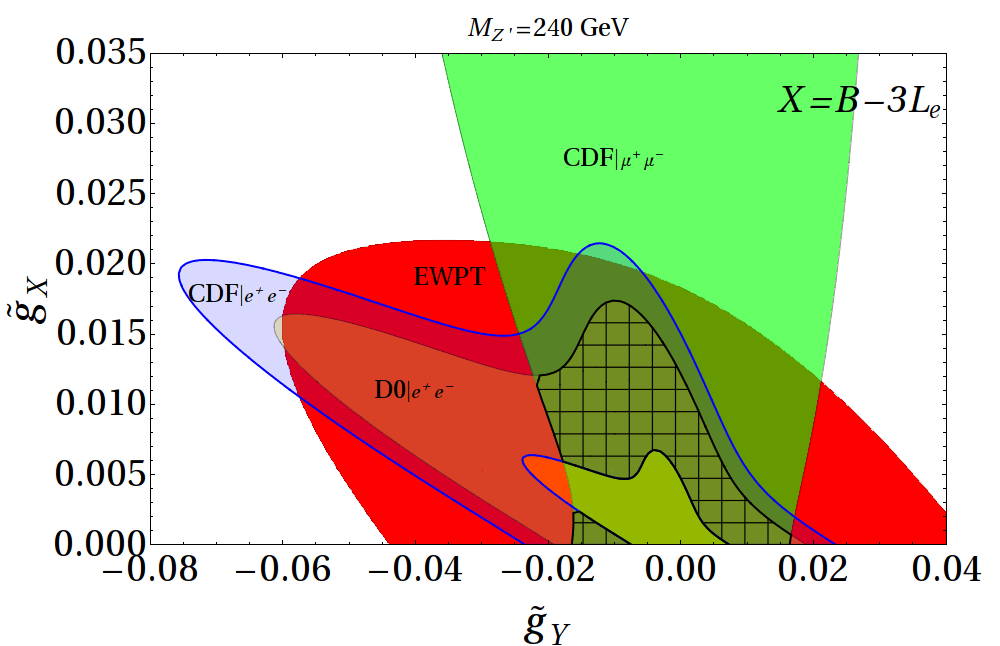}
\caption{Electrophilic model, $X$$=$$B$$-$$3L_e$. We assume $M_{Z'} = 240\,{\rm GeV}$, such that the CDF $e^+e^-$ excess can be reproduced within the blue region at 95\% CL. The other shaded regions are allowed at the same CL from precision tests (red), CDF dimuon searches (green), D0 dielectron searches (yellow, hiding behind the blue region). The intersection gives the common allowed region: it is shown in textured gray and is non-empty; hence the CDF excess is compatible with all bounds.}
\label{fig:cdf240}
}

The main peculiarity of the electrophilic model with $X=B-3L_e$ is the asymmetry between the electron and muon channels at hadronic colliders. This asymmetry is able to produce sharp signals in one channel while hiding completely in the other. We illustrate this interesting feature using the recent excess \cite{CDFepem} observed by the CDF collaboration in the dielectron spectrum, at an invariant mass near 240~GeV, which however was not observed in the dimuon channel \cite{CDFmumu}.  The question we want to answer here is the following: \emph{can such an excess be explained with a narrow $Z'$ resonance, compatibly with the negative results from the dimuon channel and with the constraints from EWPT and from flavor physics?}

The excess would correspond to a resonance with mass around 240~GeV and despite the $2.5\sigma$ significance (which amounts to more than $3\sigma$ considering only that single bin), the coupling to the SM fermions should be small ($g_{Z'} \lesssim 10^{-2}$). Since the candidate resonance mass is known, we can plot all the different bounds from EWPT and Tevatron searches together in the $(\tilde g_Y, \tilde g_x)$ plane: the result is shown in Fig.~\ref{fig:cdf240}, which displays the allowed regions from EWPT (red), CDF $e^+ e^-$ (blue) and $\mu^+ \mu^-$ (green) searches, D0 $e^+ e^-$ (yellow) searches~\cite{D0epem}.  In view of the $2.5 \sigma$ excess, the CDF $e^+ e^-$ region does not include the origin (i.e.\ the SM limit). Notice also that there is a non-trivial common region of couplings (textured in gray) which is not ruled out by any experiment and is compatible with the observed excess. For such a low-energy resonance, the first LHC runs will not be very competitive with respect to the higher luminosity data collected at the Tevatron. The values of the allowed couplings are quite smaller than those preferred by unification, which suggests that the would-be $Z'$ will hardly unify with the other SM gauge coupling at higher energies. It is however remarkable that it is possible to fit such an `exotic' flavor-breaking excess with a quite simple, minimal, renormalizable and FCNC-free new physics model.

\subsection{Muonphilic model: \boldmath{$X = B - 3 L_\mu$}} 
\label{muonphilic}

The second type of $Z'$ model we consider [corresponding to $\lambda_\mu=3$, $\lambda_e = \lambda_\tau = 0$ and $\beta = 1$ in Eq.~(\ref{xdef})] is associated to the gauging of a linear combination of $X = B - 3 L_\mu$ and $Y$. This model shares many properties with the previous one for what is related to unification and to the signatures at hadron colliders (after the obvious exchange of the electron and muon channels). On the other hand, the possibility of having a suppression or even the complete cancellation of the couplings to electrons alters completely the bounds from EWPT, making them very weak. As we will see in a moment, this feature may turn out to be crucial for the early LHC studies. 

For what concerns unification, the story is exactly the same as in the previous case: the beta functions are indeed identical, since we just exchanged the r\^ole of the first and second lepton generation. The plot of the favored region in the $(\widetilde g_Y,\widetilde g_X)$ plane is thus the same as before  and is displayed on the left-hand side of Fig.~\ref{fig:gut-ewptBLm}.
\FIGURE[t] {
\includegraphics[width=0.49\textwidth]{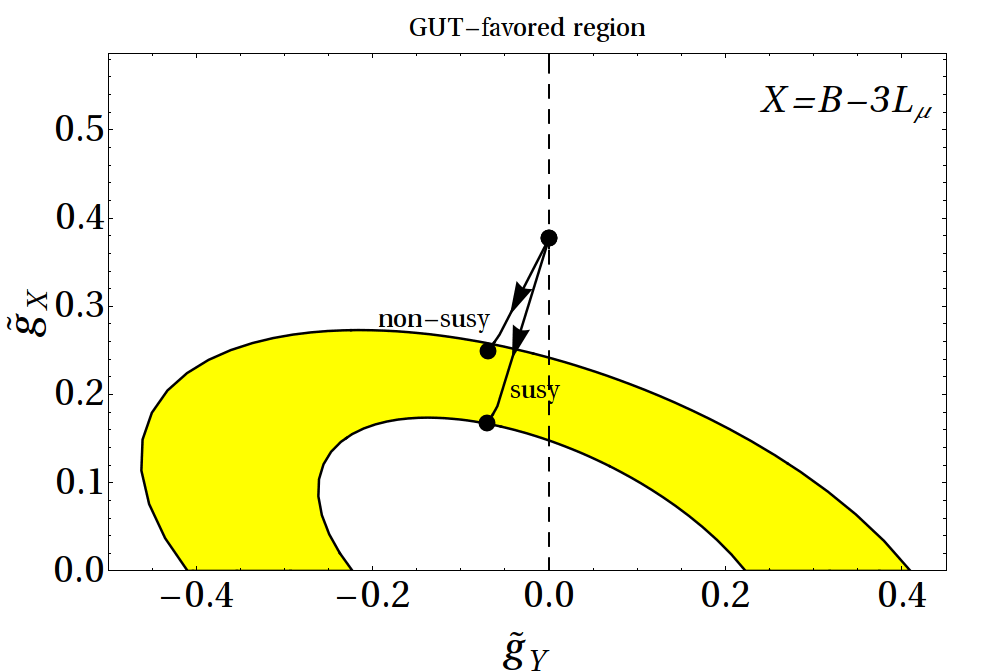}
\includegraphics[width=0.49\textwidth]{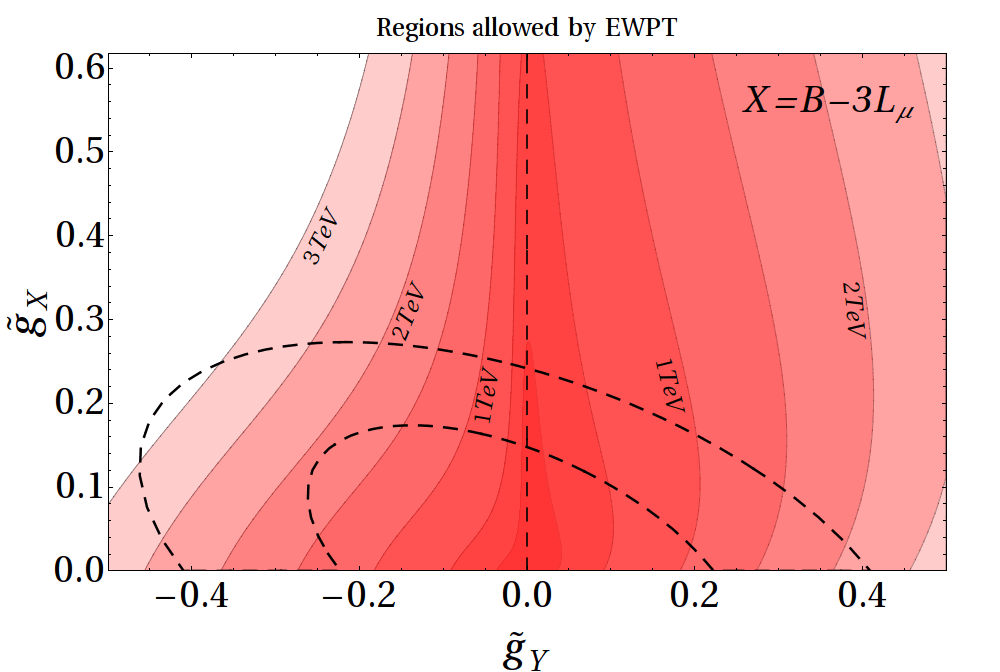}
\caption{Muonphilic model, $X$$=$$B$$-$$3L_\mu$. \emph{Left:} GUT-favored region in the $(\widetilde g_X, \widetilde g_Y)$ plane and the effect from the RGE running with boundary conditions $\alpha_U(10^{16}~{\rm GeV})=1/24$ and $\tilde g_Y(10^{16}~{\rm GeV})=0$ for the two cases (i) and (ii) discussed in the text (higher and lower black curves). \emph{Right:} The allowed regions at 95\% CL from EWPT for $M_{Z'}=200$, $500$, $1000$, $1500$, $2000$, $2500$ and $3000$~GeV (from inner to outer).}
\label{fig:gut-ewptBLm}}
On the right-hand side of the same figure, also the bounds from EWPT for different $Z'$ masses are shown. The plot is now completely different from those for $X=B-L$ and $X=B-3L_e$. First of all, there is almost no bound on the $Z'$ masses along the $\widetilde g_Y=0$ axis: these are indeed pure $B-3L_\mu$ models, which do not couple to electrons (thus no bounds from LEP2) and do not mix with the SM $Z$ boson (thus no constraints from LEP1 and APV). The only constraints come from $\nu_\mu$ DIS experiments, such as NuTeV, and from $(g-2)_\mu$, but the bounds are very weak. Notice however that pure $B-3L_\mu$ models are not stable under RGE and in general some mixing with the $Y$ will be produced; still, the region around $\widetilde g_Y=0$ is weakly constrained by EWPT, which makes hadronic colliders more sensitive to this particular type of models. 

The bounds from the Tevatron experiments are reported in Fig.~\ref{fig:tev3BLm} and are pretty similar to those of the previous section, with dielectron and dimuon channels exchanged. Notice that now the muon channel is the most powerful for probing the $\widetilde g_Y\simeq0$ region, where EWPT are weaker. Again, the plots for the LHC share the same shapes as those from the Tevatron, the size being different and dependent on the energies and luminosities considered.
\FIGURE[t]{
\includegraphics[width=0.49\textwidth]{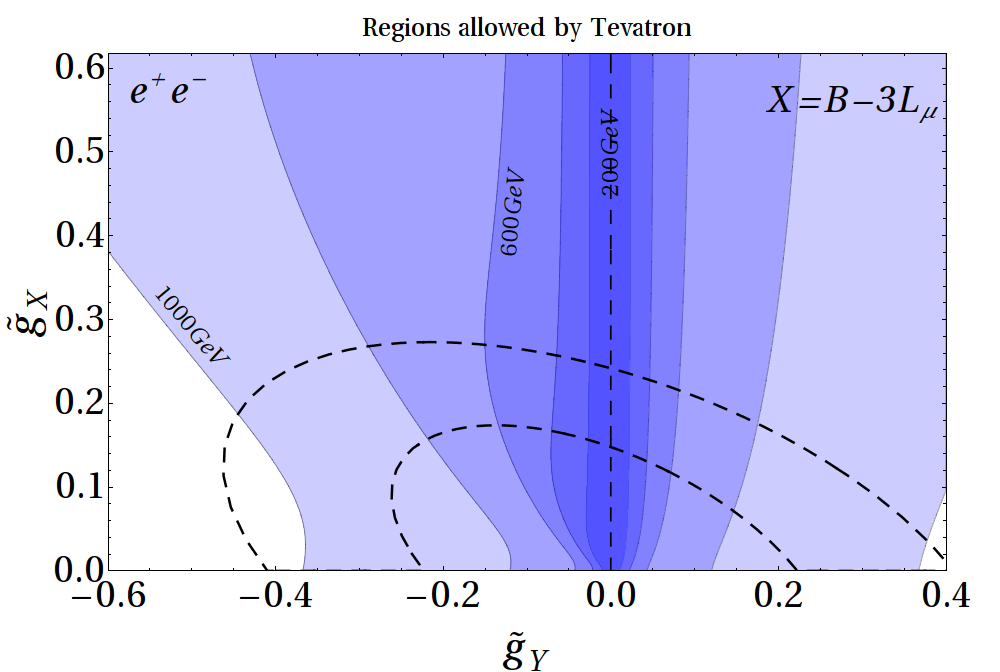}
\includegraphics[width=0.49\textwidth]{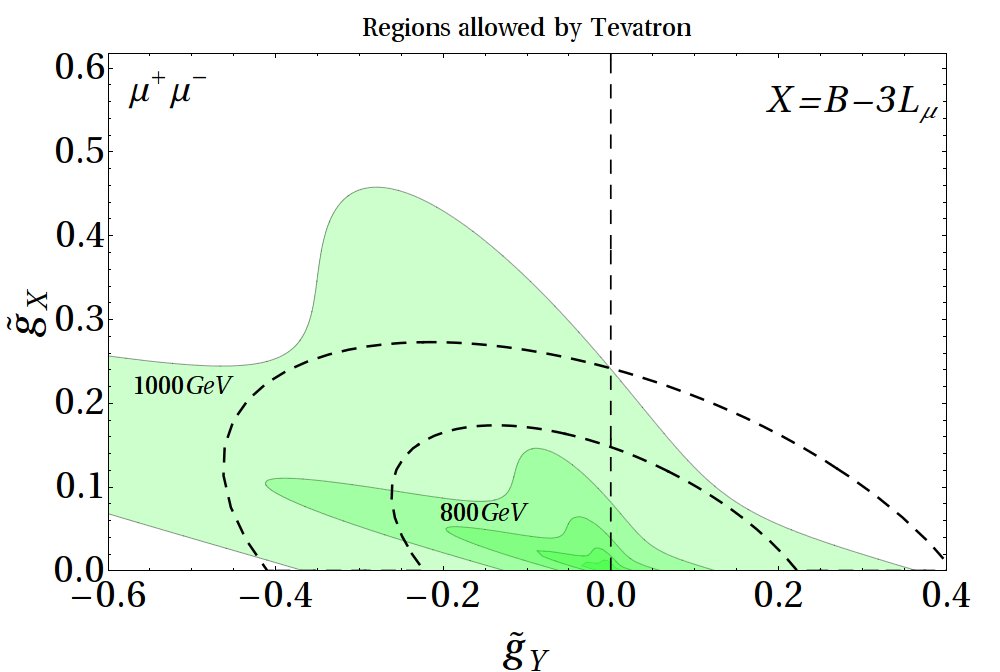}
\caption{Muonphilic model, $X$$=$$B$$-$$3L_\mu$. Allowed regions at 95\% CL from dilepton (\emph{left}) and dimuon (\emph{right}) searches at Tevatron for  $M_{Z'}=200$, $400$, $600$, $800$ and $1000$~GeV (from inner to outer).}
\label{fig:tev3BLm}}

The attractive feature of this model is thus evident. We have seen that there is a region of couplings where EWPT are particularly weak ($\widetilde g_Y\simeq0$) and where we may expect the dimuon searches at hadron colliders to be very powerful instead. We also know that for resonances above 1~TeV the higher energy of the LHC can easily outperform the Tevatron searches and bounds, even after collecting a very small integrated luminosity.  There are then regions allowed by EWPT and Tevatron where the LHC can make a discovery after collecting only few pb$^{-1}$ of data. This possibility is illustrated in Fig.~\ref{fig:lhcBLm}, which shows the combination of bounds from EWPT, Tevatron bounds and the LHC reach for $M_{Z'}=800$, 1000 and 1200~GeV. 
\FIGURE[t]{
\includegraphics[width=0.52\textwidth]{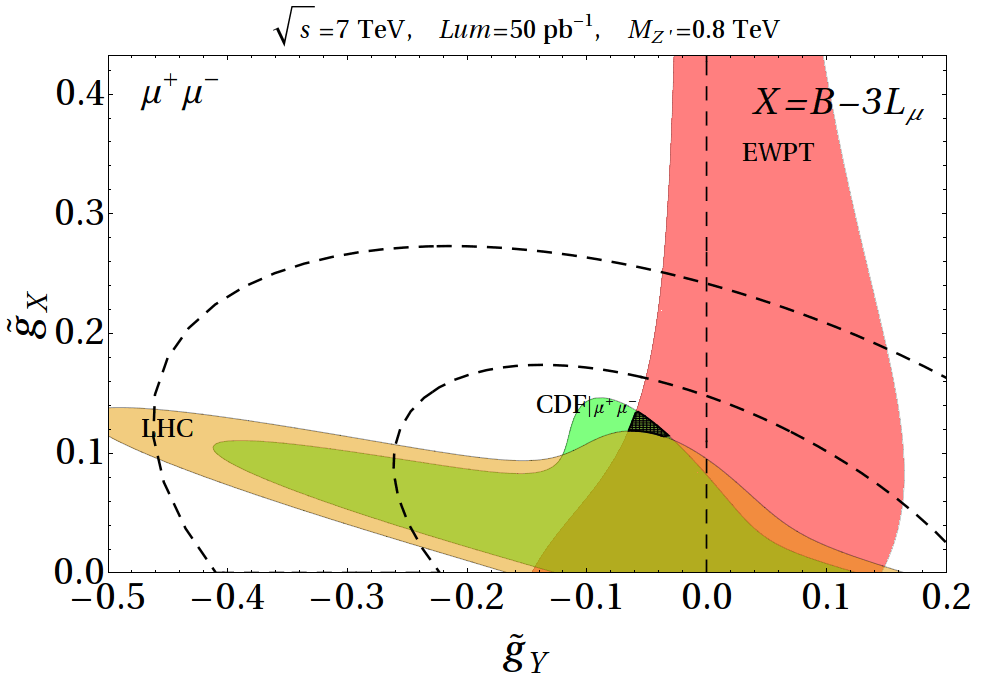}
\includegraphics[width=0.52\textwidth]{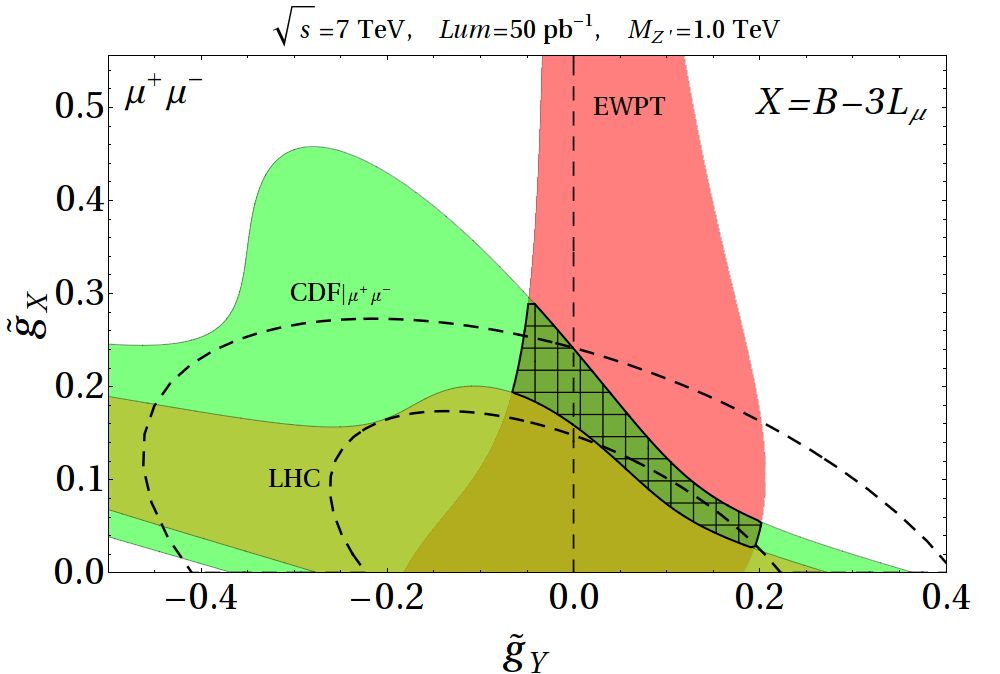}
\includegraphics[width=0.52\textwidth]{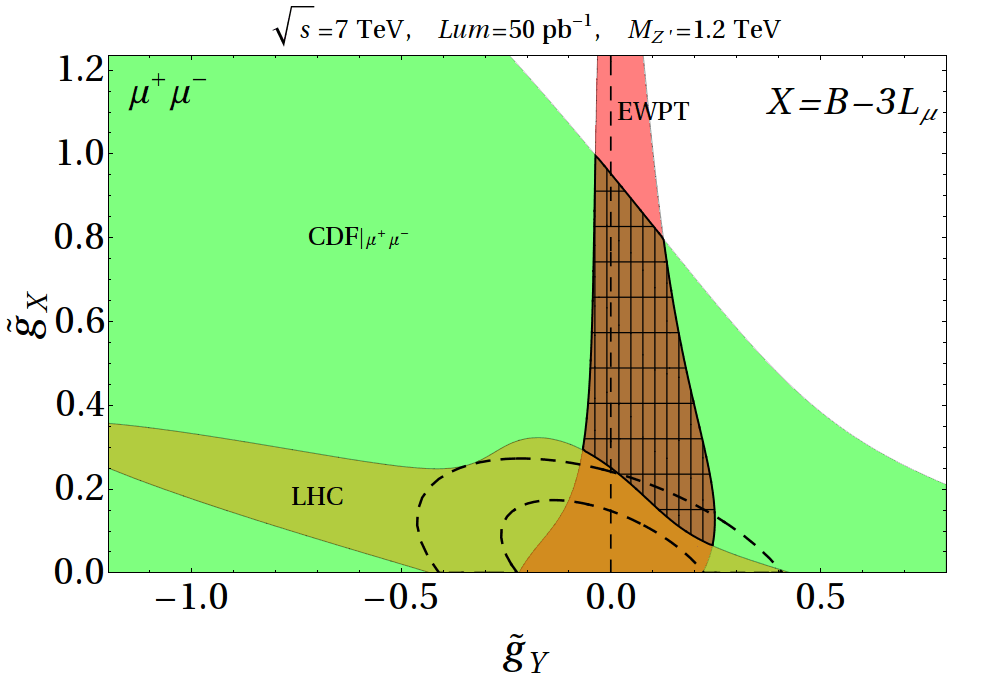}
\caption{Muonphilic model, $X$$=$$B$$-$$3L_\mu$. Regions allowed at 95\% C.L.\ by precision tests (red), Tevatron  $e^+ e^-$ data,  Tevatron $\mu^+\mu^-$ data (green)  compared to the region not accessible at $5\sigma$ level by the LHC dimuon searches (yellow) with $\sqrt{s}=7$~TeV, 50~pb$^{-1}$ of integrated luminosity. 
We assume $M_{Z'}=$800 (upper plot), 1000 (medium) and 1200~GeV (lower). The allowed region where the LHC can discover a $5\sigma$ signal is textured.}
\label{fig:lhcBLm}}
The textured regions correspond to $Z'$ models not excluded by existing experimental
bounds and accessible for a 5$\sigma$ discovery, already in the first phase of LHC: as representative values, we take 50~pb$^{-1}$ of integrated luminosity at $\sqrt{s}= 7 $~TeV. Notice that for $M_{Z'}=1$~TeV the accessible region would actually include the one preferred by unification (even considering the mixing induced by the RGE running). In the language of \cite{supmod}, these models can be considered to be \emph{supermodels}. This very simple possibility was not considered in \cite{supmod} because of possible issues with FCNC, but in Sect.~\ref{neum} we showed that such flavor problem is automatically solved in our class of models.

\FIGURE[t] {
\includegraphics[width=0.49\textwidth]{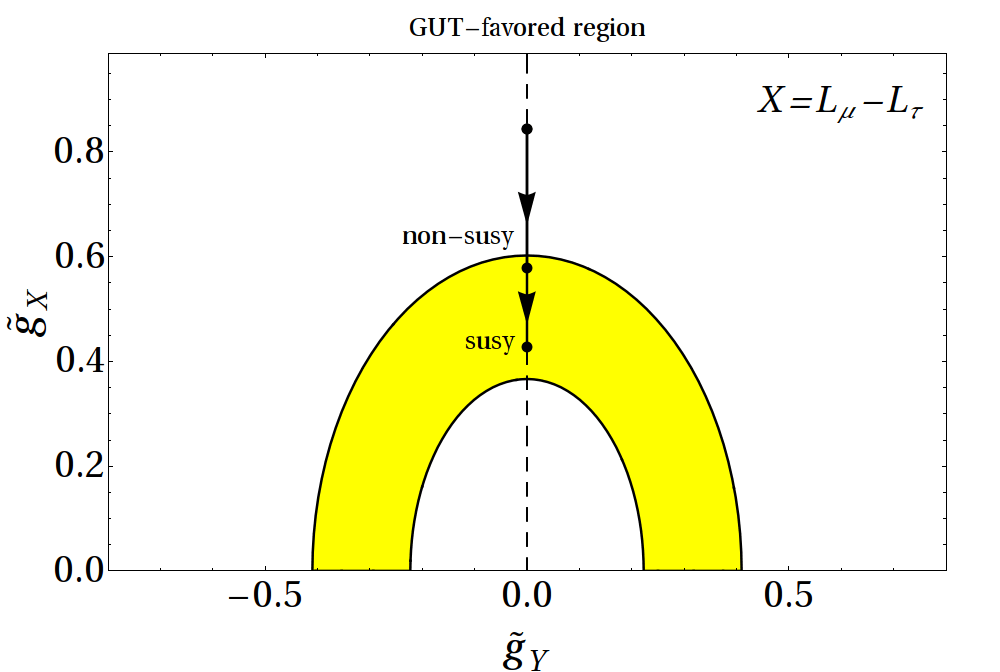}
\includegraphics[width=0.49\textwidth]{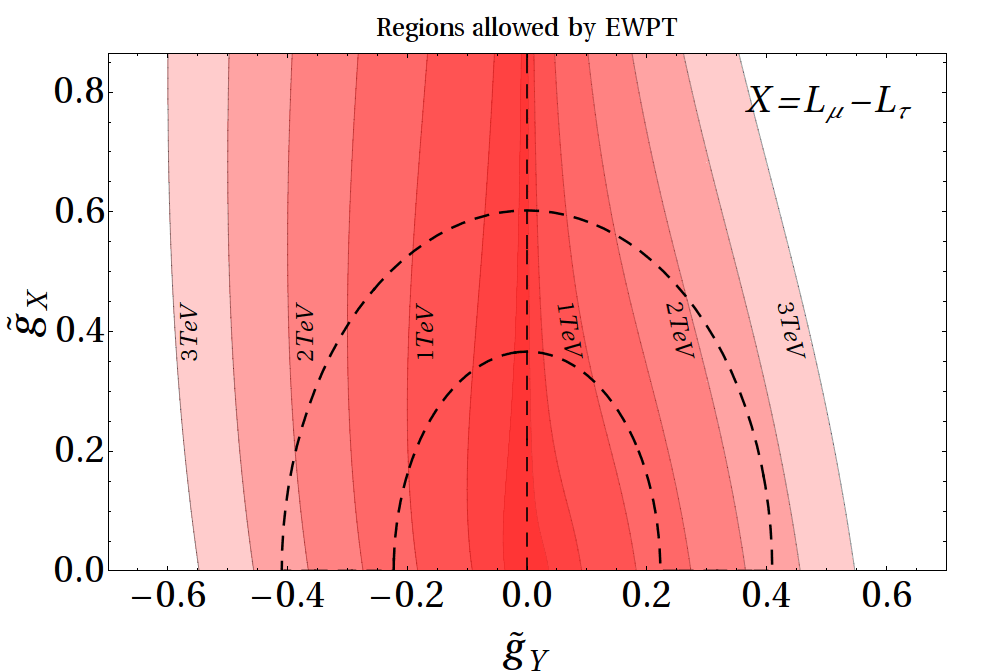}
\caption{Hadrophobic model, $X$$=$$L_\mu$$-$$L_\tau$. \emph{Left:} GUT-favored region in the $(\widetilde g_X, \widetilde g_Y)$ plane and the RGE running from  $\alpha_U(10^{16}~{\rm GeV})=1/24$ and $\tilde g_Y(10^{16}~{\rm GeV})=0$ for the two models (i) and (ii) discussed in the text (higher and lower black curves). \emph{Right:} Regions allowed at 95\% CL by precision tests for $M_{Z'}=200$, $500$, $1000$, $1500$, $2000$, $2500$ and $3000$~GeV (from inner to outer).}
\label{fig:gut-ewptLmLt}}

\subsection{Hadrophobic model:  \boldmath{$X = L_\mu - L_\tau$}} 
\label{hadrophobic}

Finally, the $B-3L_\tau$ model does not lead to clean peaks in the lepton invariant masses at hadron colliders. It is weakly constrained by hadronic  Tevatron data and by the non-observation of matter effects in $\nu_\mu\leftrightarrow\nu_\tau$ atmospheric oscillations. Therefore we prefer to consider a different last example  [corresponding to $\lambda_\tau=-\lambda_\mu=1$, $\lambda_e = 0$ and $\beta = 0$ in Eq.~(\ref{xdef})], where the $Z'$ couples to a linear combination of $Y$ and $X = L_\mu - L_\tau$. The peculiarity of this model is that, in the pure $L_\mu-L_\tau$ phase, the $Z'$ couples neither to electrons nor to quarks, hiding completely from EWPT and direct collider searches. In this particular case the $Z'$ can be extremely light, the only bounds coming from the muon magnetic moment. When the mixing with $Y$ is considered, the $Z'$ couplings to quarks and electrons switch on and the usual bounds apply again, albeit suppressed by the mixing angle $\theta^\prime$ when the latter is small. Unlike the previous cases, we thus expect a dark region along the $\widetilde g_Y=0$ axis, blind both to bounds and to discoveries at colliders.

Also the properties under RGE running are different in this model. Indeed, no kinetic mixing is generated at one loop if it vanishes at tree level:
\begin{center}
\begin{tabular} { l | c   c   c  }
model & $b_{YY}$ &$b_{YX}$ &$b_{XX}$ \\
\hline
\phantom{i}i) non-susy  & 41/6 & 0 & 7\\
ii) susy & 11 & 0  & 18
\end{tabular}
\end{center}
The absence of kinetic mixing makes the axis $\widetilde g_Y=0$ an attractor under the RGE evolution, which means that pure $L_\mu-L_\tau$ models are stable under quantum effects and do not mix appreciably with $Y$, even in the case when small threshold corrections are present at the GUT scale. This can be seen by looking at the RGE for $g_Y$ near $g_Y=0$:
\begin{equation}
\frac{dg_{Y}}{dt} = \frac{g_Y}{16 \pi^2}  \left( 2b_{YY} {g'}^{2}+ b_{XX} g_{X}^{2} + {\cal O} (g_Y^2) \right) \, ,
\end{equation}
which shows that $dg_{Y}/dt$ vanishes at $g_Y=0$ and has the same sign as $g_Y$, so that the latter flows to zero at low energies. The GUT-favored region in this case is represented in Fig.~\ref{fig:gut-ewptLmLt}, the absence of kinetic mixing makes the region symmetric around the vertical axis.

The constraints from EWPT at various masses are displayed in the same Fig.~\ref{fig:gut-ewptLmLt}. As expected, no useful constraints are present for a $Z'$ associated to pure $L_\mu-L_\tau$. Away from the vertical axis the constraints are basically on the `$Y$-amount' $\widetilde g_Y$ of the $Z'$, almost independently of $\widetilde g_X$.

The bounds from the Tevatron searches are plotted in Fig.~\ref{fig:tev3LmLt}. 
In the dielectron channel, the region around the unconstrained $\widetilde g_Y=0$ line
(pure $L_\mu-L_\tau$) becomes larger with increasing $g_X$, because  larger $g_X$ increases the couplings to muons and taus, thus dumping the branching ratio to electrons.

\FIGURE[t]{
\includegraphics[width=0.49\textwidth]{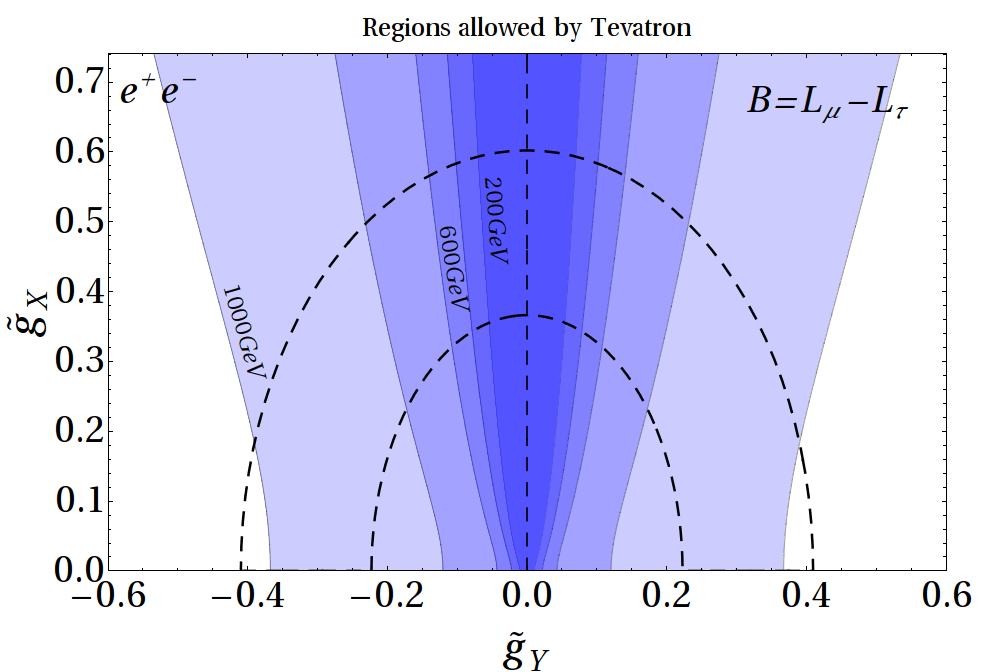}
\includegraphics[width=0.49\textwidth]{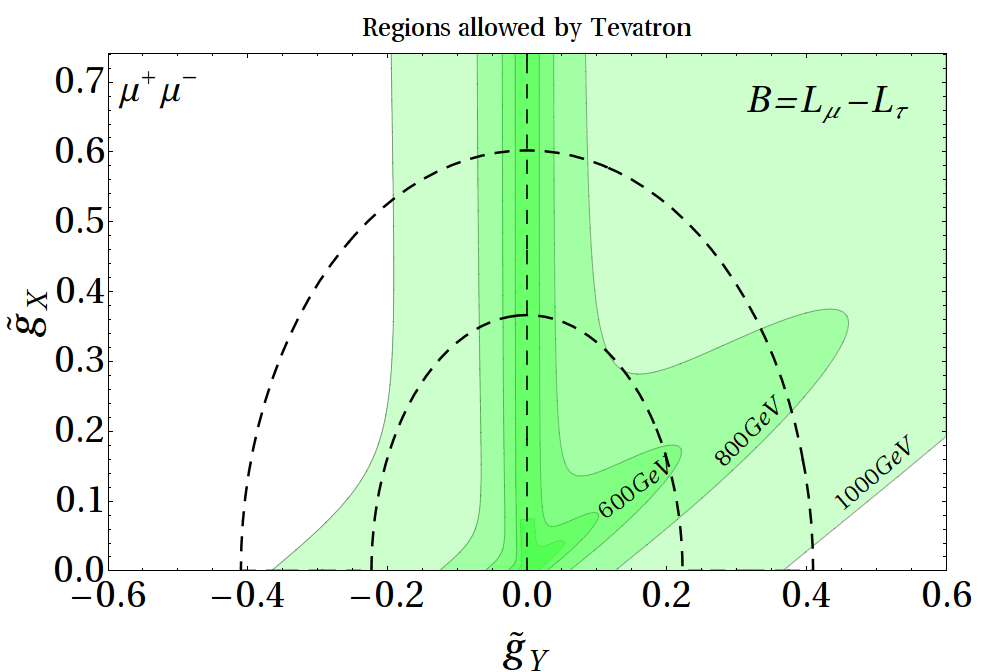}
\caption{Hadrophobic model, $X$$=$$L_\mu$$-$$L_\tau$. Regions allowed at 95\% CL by dilepton (\emph{left}) and dimuon (\emph{right}) searches at Tevatron for  $M_{Z'}=200$, $400$, $600$, $800$ and $1000$~GeV (from inner to outer).}
\label{fig:tev3LmLt}}
\FIGURE[t]{
\includegraphics[width=0.49\textwidth]{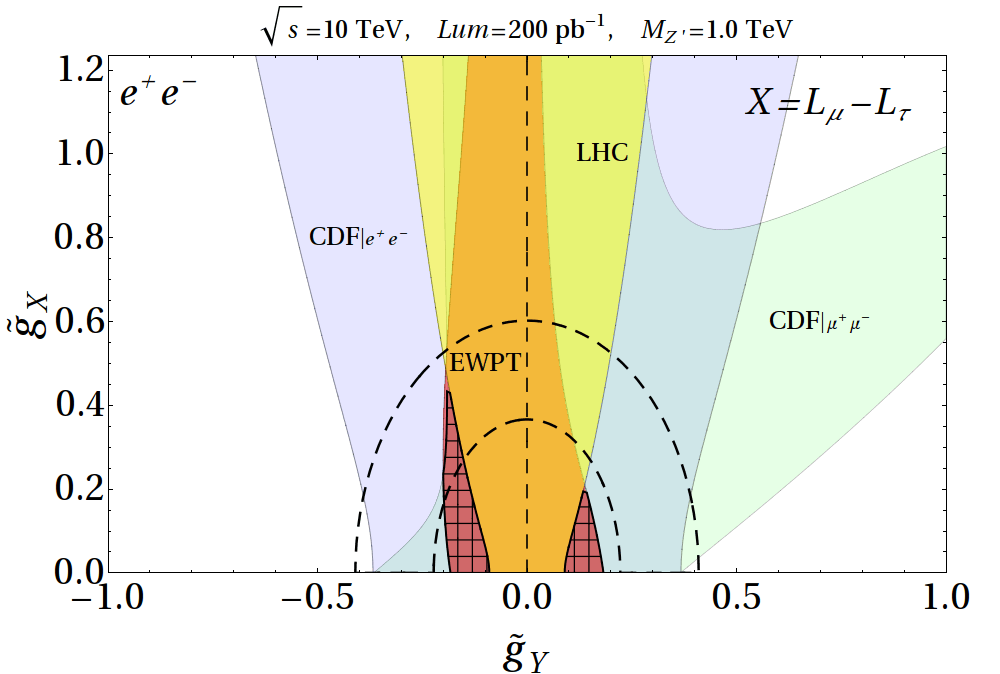}
\includegraphics[width=0.49\textwidth]{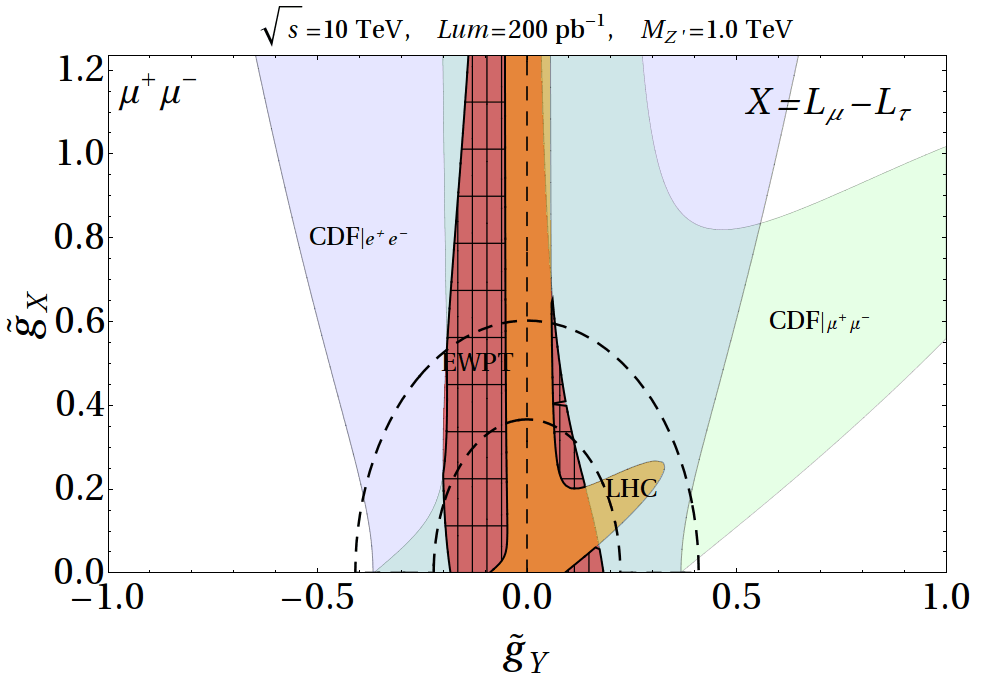}
\caption{Hadrophobic model, $X$$=$$L_\mu$$-$$L_\tau$. Comparison of 95\% CL bounds from EWPT (red) and Tevatron (blue for $e^+ e^-$ and green for $\mu^+\mu^-$) and discovery reach at the LHC (yellow) with $\sqrt{s}=10$~TeV, 200~pb$^{-1}$ of integrated luminosity for $M_{Z'}=$1~TeV for dielectron and dimuon searches respectively. The textured region is the one accessible to a $5\sigma$ discovery and compatible with existing bounds.}
\label{fig:lhcLmLt}}

The dimuon bounds are not symmetric under $\tilde{g}_Y\to -\tilde{g}_Y$, and are less constraining for $\widetilde g_X\sim \widetilde g_Y$, because of a partial cancellation between the $Y$ and $L_\mu-L_\tau$ components of the $Z'$ couplings to muons.

Coming to the LHC discovery reach, there are regions accessible in the early phase, both in the $e^+ e^-$ and $\mu^+ \mu^-$ channels. These are illustrated in Fig.~\ref{fig:lhcLmLt}, where we plot, as a representative example, the accessible regions for a $5\sigma$ discovery at the LHC after 200~pb$^{-1}$ at $\sqrt{s} =10$~TeV, for $M_{Z'}=1$~TeV. The muon channel seems quite powerful, exploring a sizable region of the $(\widetilde g_Y,\widetilde g_X)$ plane for these masses, except of course the dark region around $\widetilde g_Y=0$. For smaller masses the Tevatron bounds start dominating, while at larger masses the bounds from EWPT eventually overcome the early LHC reach.

\section{Conclusions}
\label{concl}
 
Starting from the SM augmented by three right-handed neutrinos, to account for neutrino masses, we explored the most general extra $U(1)$ that can be gauged compatibly with the following requirements: the theory must be renormalizable, anomaly-free, compatible with the observed fermion masses and mixings and with the absence of extra violations of flavor, beyond the SM CKM matrix and those induced in the lepton sector by the light neutrino masses. The most general $Z'$ boson of this kind was found to be a linear combination of the hypercharge $Y$, $B-3L_e$, $B-3L_\mu$, $B-3L_\tau$. Although this statement is not new, we showed that it still holds now that neutrino masses and mixings have been observed. Unlike in the quark sector, where individual baryon flavors cannot be gauged, this is still possible in the lepton sector, as the resulting flavor-changing neutral currents are suppressed by the small neutrino masses, as in the SM.

A new $Z'$ gives clean easy signals at the LHC but also in previous experiments, thereby we computed the allowed parameter space comparing the present bounds from EWPT and the Tevatron to the LHC capabilities. It is important to keep in mind that a given $Z'$, associated with a definite direction in the four-dimensional space defined by $Y$ and $B-3L_a$ ($a=e,\mu,\tau$), is specified by two parameters: its mass $M_{Z'}$ and its effective gauge coupling $g_{Z'}$. The often-reported limits on the $Z'$ mass \cite{reviews} tacitly assume some specific arbitrary value of $g_{Z'}$. Although we identified the ranges of $g_{Z'}$ suggested by grand unification, making use of the appropriate RGE from the GUT scale to the weak scale, this is just a theoretical hypothesis. Both parameters must be kept in a complete phenomenological analysis, and both are important. The left-hand side of Fig.~\ref{fig:Mg} shows a typical case: the region where the early LHC can outperform all previous experiments corresponds to $M_{Z'} \circa{>} 500 \, {\rm GeV}$ and $g_ {Z'}$ significantly smaller than the GUT-favored values. Indeed, if the $Z'$ is light, $\sqrt{s}$ is not important, and the early LHC cannot compete with the $\sim50$ times larger luminosity already accumulated at the Tevatron. Furthermore, too large values of $g_{Z'}/M_{Z'}$ have already been excluded indirectly by precision data.
\FIGURE[t]{
\includegraphics[width=0.49\textwidth]{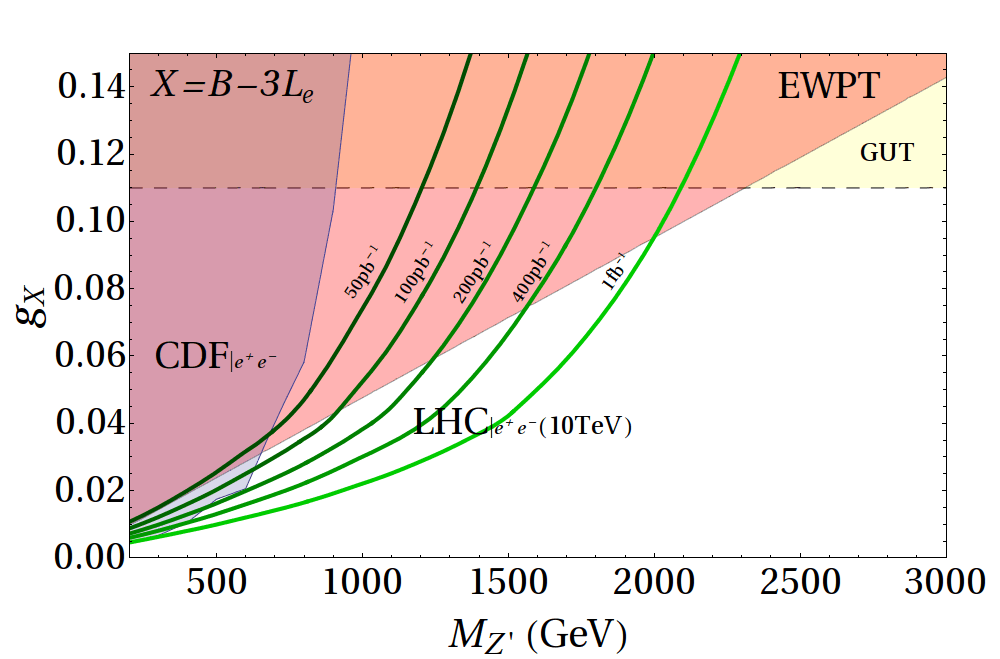}
\includegraphics[width=0.49\textwidth]{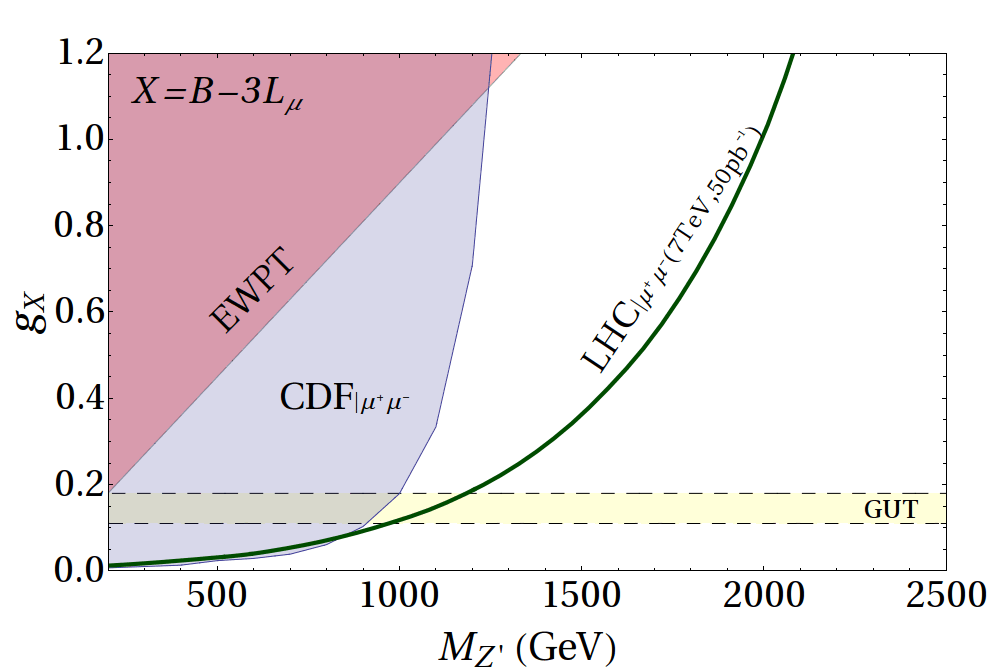}
\caption{Early LHC $5\sigma$ discovery reach compared with bounds from
EWPT and Tevatron. Left: pure $B-3L_e$. Right: pure $B-3L_\mu$. For values of the LHC energy and luminosity higher than those indicated in the figure, the sensitivity in $g_X$ roughly scales as (luminosity)$^{-1/2}$, while the sensitivity in $M_{Z'}$ roughly scales as $\sqrt{s}$.}
\label{fig:Mg}}

\bigskip 

We focussed on the following representative possibilities.
\begin{enumerate}
\item 
The electrophilic $B-3L_e$, possibly mixed with the hypercharge $Y$. The left-hand side of Fig.~\ref{fig:Mg} compares the LHC sensitivity with the bounds from Tevatron and from EWPT, it has the typical shape of similar graphs for generic $Z'$ models. The main novelty with respect to the $B-L$ case studied in~\cite{svz} is the possibility to fit the CDF excess in the $e^+e^-$ channel with $M_{Z'}  \approx 240$ GeV compatibly with all other constraints, in the region of parameter space shown in Fig.~\ref{fig:cdf240}. Being light, such $Z'$ will be tested at the Tevatron with more luminosity sooner than at the LHC.
\item
The muonphilic $B-3L_\mu$, possibly mixed with the hypercharge $Y$. The main novelty is that this possibility is very weakly constrained by precision data when the mixing with $Y$ is small: the two experiments most sensitive to the pure $B-3L_\mu$ case, $g-2$ and NuTeV, show anomalies, but it is not easy to fit them and the impact on the allowed parameter space is negligible. As illustrated on the right-hand side of Fig.~\ref{fig:Mg}, in the case of pure $B-3L_\mu$, a significant region of parameter space can therefore be explored even by the initial stage of the LHC with reduced energy and luminosity.
\item
Finally, the hadrophobic $L_\mu-L_\tau$, possibly mixed with the hypercharge $Y$, can give detectable signals at the LHC, but not easily in the early phase. Unlike the first two cases, mixing with hypercharge
is not automatically generated by the RGE running.
\end{enumerate}

\acknowledgments
We thank G.~Altarelli, T.~Dorigo, F.~Feruglio, G.~Polesello and E.~Torassa for discussions. This research was supported in part by the European Programme {\em Unification in the LHC Era},  contract PITN-GA-2009-237920 (UNILHC), and by the Fondazione Cariparo Excellence Grant {\em String-derived supergravities with branes and fluxes and their phenomenological implications}.

%

%
\end{document}